\documentclass[preprint,12pt]{elsarticle}

\usepackage{longtable}

\usepackage{wrapfig}
\usepackage{xcolor}
\usepackage{pdfpages}
\usepackage{pdflscape}
\usepackage{geometry}
\usepackage{tabularx}
\usepackage{booktabs}
\usepackage{float}
\usepackage{enumitem}
\usepackage{needspace}
\usepackage{pdflscape}
\usepackage{amssymb}
\usepackage[many]{tcolorbox}
\usepackage{url}
\usepackage{hyperref}
\biboptions{sort&compress}

\newtcolorbox{fancybox}[1][]{
  enhanced,
  attach boxed title to top center={yshift=-3mm,yshifttext=-1mm},
  colback=blue!5!white,
  colframe=blue!25!black,
  colbacktitle=blue!25!black,
  fonttitle=\bfseries,
  title=#1,
  boxed title style={size=small, colframe=blue!45!black, colback=blue!45!white},
  drop fuzzy shadow,
  width=\textwidth,
  breakable=true
}

\AtBeginDocument{%
  }

\begin{document}

\title{Ethical Concerns of Generative AI and Mitigation Strategies: A Systematic Mapping Study}

\author[1]{Yutan Huang}
\author[1]{Chetan Arora}
\author[1]{Wen Cheng Huong}
\author[2]{Tanjila Kanij}
\author[3]{Anuradha Madulgalla}
\author[1]{John Grundy}

\affiliation[1]{organization={Faculty of Information Technology, Monash University},
            city={Clayton}, 
            state={Victoria},
            country={Australia}}

\affiliation[2]{organization={School of Science, Computing and Engineering Technologies, Swinburne University},
            city={Hawthorn}, 
            state={Victoria},
            country={Australia}}

\affiliation[3]{organization={School of Information Technology, Deakin University},
            city={Geelong}, 
            state={Victoria},
            country={Australia}}

\begin{abstract}
  Generative AI technologies, particularly Large Language Models (LLMs), have transformed numerous domains by enhancing convenience and efficiency in information retrieval, content generation, and decision-making processes. However, deploying LLMs also presents diverse ethical challenges, and their mitigation strategies remain complex and domain-dependent. This paper aims to identify and categorize the key ethical concerns associated with using LLMs, examine existing mitigation strategies, and assess the outstanding challenges in implementing these strategies across various domains. We conducted a systematic mapping study, reviewing 39 studies that discuss ethical concerns and mitigation strategies related to LLMs. We analyzed these ethical concerns using five ethical dimensions that we extracted based on various existing guidelines, frameworks, and an analysis of the mitigation strategies and implementation challenges. Our findings reveal that ethical concerns in LLMs are multi-dimensional and context-dependent. While proposed mitigation strategies address some of these concerns, significant challenges still remain. Our results highlight that ethical issues often hinder the practical implementation of the mitigation strategies, particularly in high-stake areas like healthcare and public governance; existing frameworks often lack adaptability, failing to accommodate evolving societal expectations and diverse contexts.
\end{abstract}

\begin{keyword}

Generative AI
\sep AI Ethics
\sep Large Language Models (LLMs)
\sep Systematic Mapping Study

\end{keyword}

\maketitle

\section{Introduction}~\label{sec:introduction}

The evolution of Generative AI, particularly Large Language Models (LLMs), has seen remarkable advancements since 2020 with the introduction of models like ChatGPT and Bard. LLMs have revolutionized tasks, such as writing assistance, code generation, and customer support automation, by leveraging vast amounts of data to generate coherent and contextually relevant natural language (NL) responses~\cite{alawida2023comprehensive, arora2024advancing}. As a subset of Generative AI—systems designed to create new content—LLMs go beyond traditional AI techniques, which focus primarily on analyzing existing data. LLMs, in contrast, are capable of generating text, images, and music that mimic human creativity~\cite{linkon2024advancements}. This capability is powered by advancements in neural network architectures, especially transformers, which enable LLMs to learn the nuances of human language and produce semantically accurate content~\cite{hagos2024recent}.

LLMs, such as OpenAI's GPT-4 and Google's Gemini, utilize transformer architecture to understand and generate human-like text. GPT-4, for instance, employs a transformer-decoder architecture with billions of parameters, allowing it to generate detailed and contextually relevant text across various topics \cite{bengesi2024advancements}. 

LLMs have been used to address numerous challenges, offering innovative solutions across sectors such as healthcare, education, and finance. It has the potential to bring efficiency and effectiveness into these areas~\cite{reddy2024generative}. However, as a rapidly emerging technology, LLMs currently operate with limited regulations or oversight, which raises significant ethical concerns~\cite{bautista2024ethical}. Without mature and robust guidelines,  these potent tools risk being misused or improperly applied. For instance, there have been cases where AI-generated content has perpetuated biases or disseminated misinformation, underscoring the critical need for comprehensive ethical guidelines \cite{bontridder2021role, GermaniFederico2024TDNo,SanchezThomasW.2024TECo,rezaei2024fairness}. Although ethics is a broad concept that varies across contexts and fields, understanding the ethics around LLMs is becoming increasingly important for guiding the responsible use and development of these technologies.

To effectively mitigate these risks and harness the full potential of LLMs, it is crucial to understand the existing ethical concerns and strategies proposed to address these issues. To this end, we present our systematic mapping study (SMS) aimed at identifying primary studies that have investigated the various key ethical challenges associated with LLMs usage and by analyzing these studies to provide insights into developing a more comprehensive ethical framework to guide their responsible use. We observed that all the articles provided their understanding of the ethical concerns and categorized these concerns into specific ethical dimensions. The ethical dimensions were categorized based on existing ethical guidelines and regulatory frameworks. These ethical dimensions refer to topics that group similar ethical issues arising within the field, e.g., privacy and bias. We also found that while all the articles proposed different strategies to address these ethical concerns, most were conceptual ideas with a noticeable lack of evaluation regarding their effectiveness in resolving ethical issues using generative AI.

We followed Kitchenham et al.’s~\cite{kitchenham2022segress} {six-step protocol for systematic reviews: defining research questions, developing a search strategy, selecting studies with clear inclusion and exclusion criteria, assessing study quality, extracting data, and synthesizing results} for performing systematic reviews. \textcolor{black}{In parallel, we adopted} Peterson et al.’s~\cite{petersen2015guidelines}  \textcolor{black}{systematic mapping guidelines, which emphasize designing a classification schema, mapping publications into that and identifying research trends and gaps}. 

\textcolor{black}{We note that several recent reviews have explored AI ethics in different domains. However, none of them cover a cross-domain, methodologically mapped approach that includes both peer-reviewed studies and industry/governmental guidelines, as well as the empirical evaluation status of the proposed mitigation strategy and implementation challenges. For example, Li et al. \cite{li2022ethics} systematically mapped ethical concerns and mitigation strategies for AI in healthcare, highlighting privacy, transparency, and accountability issues in clinical decision support systems; however, their scope was confined to medical settings. On the other hand, Sánchez et al.~\cite{SanchezThomasW.2024TECo} examined ethical dilemmas in AI driven urban planning, focusing only on governance and public safety, Morley et al.~\cite{morley2020initial} reviewed publicly available AI ethics tools and methods for translating high level principles into practice, yet they did not evaluate their effectiveness in real‐world settings. Atlam et al. \cite{atlam2024slm} have provided a high-level mapping of 127 ethics studies; they did not assess the empirical evaluation status or the implementation challenges of the studies they identified. By contrast, our study represents a cross-domain mapping that (a) involves both peer-reviewed studies and industry/governmental guidelines, (b) assesses the empirical evaluation outcomes of each mitigation strategy, and (c) identifies the implementation challenges of the strategies in real-world settings.} We answer the research questions (RQs) noted below in our study. \textcolor{black}{In our RQs and throughout the paper, we refer to the term \emph{ethical dimensions}. We use the term ethical dimensions to mean broad, higher-order constructs that group together closely related ethical concerns. An ethical dimension is therefore a conceptual lens, rather than a single measurable variable through which multiple concrete issues can be examined and compared. Specifically, we consider Safety, Privacy, Transparency, Bias and Accountability as the core dimensions based on our review of academic studies and existing guidelines.}

\textbf{RQ1. What are the ethical dimensions defined in the use of generative AI across various fields?} In RQ1, we wanted to identify key ethical dimensions associated with deploying and creating LLMs by analysing relevant literature. These ethical dimensions were then mapped against existing ethical guidelines and regulatory frameworks to assess whether they are acknowledged within these standards.

\textbf{RQ2. What strategies are used or proposed to address the ethical concerns of using generative AI across different fields?} Having identified specific ethical dimensions and issues in RQ1, we then examined mitigation strategies mentioned in the analyzed primary studies used to address these ethical concerns across different dimensions.

\textbf{RQ3. What are the challenges when implementing the strategies?}  We reviewed any reported challenges associated with implementing the identified mitigation strategies, detailing specific limitations and obstacles encountered in practice.

The main contributions of this mapping study include:
\begin{itemize}
\item We identified a list of 39 primary studies that focus on the ethical concerns of using generative AI. All the studies identified various ethical concerns related to specific ethical dimensions and provided strategies to address these concerns. 
\item Of these 39 papers, 13 conducted some form of empirical evaluation to either investigate or validate a proposed strategy. The remaining 26 papers are conceptual papers that propose strategies but do not conduct empirical evaluations, or in some cases, empirical evaluation is not feasible.
\item We identified the most prominent ethical dimensions where ethical concerns are concentrated and those that require more focused attention.
\item We identified a set of key research directions necessary to address ethical issues associated with the use of LLMs in practice.
\end{itemize}

The rest of the paper is structured as follows: Section 2 provides the background and related work on ethical concerns of the use of generative AI. Section 3 details our methodology for conducting a literature search and selection process on ethical considerations of generative AI. Section 4 reports the results from the selected primary studies. Section 5 discusses key results and summaries. Section 6 addresses threats to validity. Section 7 concludes our study.

\vspace{5\baselineskip}
\section{Background}~\label{sec:background}

\subsection{Terminology}~\label{subsec:Terminology}

Ethics is the discipline (or philosophy) that deals with conduct and questions of morality, exploring what is right and wrong, as well as key principles that govern human behaviors~\cite{dewey2022ethics}. This exploration includes normative ethics, which seeks to establish general principles for how people should act; applied ethics, which addresses concrete ethical issues in various fields; and metaethics, which involves analyzing the nature of moral judgments and the underlying assumptions of ethical theories~\cite{sivasubramaniam2021assisting,poff2023academic}.

AI ethics is a multidisciplinary field that examines the ethical, legal, and social implications of AI systems~\cite{jobin2019global}. It involves developing frameworks, principles, and guidelines to ensure that AI technologies are designed, implemented, and used to align with societal values, human rights, and ethical standards~\cite{anderson2021ai}. AI ethics is heavily grounded in normative ethics, as it seeks to establish principles and guidelines for designing and using AI. Normative questions in AI ethics include what constitutes fair AI, ensuring AI respects human rights, and developers' and users' responsibility in mitigating AI’s potential harms~\cite{poder2021ai}. AI ethics is also primarily a form of applied ethics, as it directly deals with practical ethical issues in real-world contexts, including the ethical deployment of AI in areas like healthcare and law enforcement, which concerns specific scenarios and case studies~\cite{kazim2021high}. Metaethical questions are also addressed in the context of AI ethics, including the debate over whether AI systems can be considered moral agents and what ethical responsibilities should be assigned to them~\cite{coeckelbergh2020ai}. \textcolor{black}{In other words, AI ethics applies ethical (moral) frameworks to considerations such as the responsibilities of developers and deployers toward affected stakeholders, the harms and benefits that should be accounted for in algorithmic decision making, the values in practice~\cite{mittelstadt2016ethics}.}

Ethical dimensions are the ethical considerations that arise when developing, deploying, and utilizing LLMs based on existing ethical guidelines and regulatory frameworks. 
When using LLMs various ethical dimensions may need to be addressed. 

Mitigation strategies are action-oriented, specific, and actionable approaches designed or proposed by the authors to address and resolve ethical issues directly; they can be evaluated. Recommendations are guidance-oriented, broader suggestions or guidelines proposed by authors that aim to influence future actions or policies regarding generative AI ethics; they are not usually subject to empirical testing or short-term evaluation. 
In our analysis of primary studies in this paper, we identify whether the strategies mentioned in the studies have been evaluated in practice. A strategy is evaluated if it has undergone some form of empirical testing, validation, or assessment to determine its effectiveness. If the strategy is theoretical or conceptual, based on logical reasoning, best practices, or expert opinion without verification of actual impact, it is not evaluated.

\subsection{Evolution and advancements of Generative AI and LLMs}

AI ethics have been discussed since the 1950s, and the discussion related to LLMs began in 2018, with various scholars and institutions proposing frameworks to address the ethical implications of AI technologies~\cite{floridi2018ai4people}. However, it was not until after 2020 that authorities began implementing concrete regulatory and policy frameworks to govern these rapidly emerging and adopted AI technologies. The increasing complexity and capability of LLMs, as well as the incidents of data breach and misuse, underscored the urgent need for robust ethical guidelines and regulatory oversight~\cite{eitel2021beyond}. For instance, the mass data leakage from Chinese applications in 2020 highlighted significant privacy concerns, raising a global call for stronger data protection measures against~\cite{hickman2021trustworthy}. Below, we review the evolution of generative AI and LLMs, the ethical concerns they raise, and the regulatory responses from major jurisdictions and industry leaders.

Initially, AI research focused on rule-based systems and basic machine learning algorithms, but the development of Generative Adversarial Networks (GANs) marked a breakthrough~\cite{goodfellow2014generative}. GANs enabled AI to create realistic images, videos, and text, which marked the foundation of Generative AI and made the space for further advancements. The introduction of the Transformer model in 2017 revolutionized Natural Language Processing (NLP), leading to the creation of powerful LLM models like GPT, which advanced the generative AI field to a large extent~\cite{vaswani2017attention}.

\subsubsection{OpenAI’s GPT Series}

These models leverage large amounts of data and to perform complex language processing tasks with proficiency. GPT-3, introduced in 2020, comprised 175 billion parameters, was well known for its size and scope, has been used across a wide range of applications, from creative writing to code generation and translation, question answering with fine-tuning, demonstrating its learning capabilities where the model can learn from a few examples of a given task~\cite{brown2020language}. 

The successor of GPT-3, GPT-4, advances these capabilities by incorporating more parameters to improve its accuracy, fluency and versatility in tasks. It is evident in areas requiring context-sensitive information, such as legal document drafting, medical diagnosis support, and scientific research summarization~\cite{alto2023modern}. There are several other LLMs. However, the GPT series of LLMs is the most widely known; hence, we only mention them, and specific LLMs are beyond the scope of this paper.

\subsection{Ethical Implications of LLMs}

The ethical concerns surrounding LLMs have been extensively discussed in the literature, with bias and privacy being primarily discussed. Bender et al.~\cite{bender2021dangers} highlighted the risks of bias and misinformation inherent in LLM models as LLMs trained on large, diverse datasets from the internet often reflect and amplify societal biases, potentially leading to harmful stereotypes and discrimination. This issue is compounded by the ``black box'' nature of LLMs, where the decision-making processes are not easily interpretable and opaque, making LLMs difficult to identify and correct biases that occurred~\cite{rudin2019stop}. Additionally, the ability of LLMs to generate realistic text can be misused for creating misinformation and conducting social engineering attacks, raising significant ethical concerns.

On the other hand, the extensive data requirements for training LLMs raise serious privacy issues. In 2017, Shokri et al.~\cite{shokri2017membership} demonstrated that AI models could be vulnerable to membership inference attacks, where an attacker can determine whether a specific individual’s data was included in the training set.  This has underscored the need for robust privacy-preserving techniques in developing and deploying LLMs. In 2021, Facebook experienced a significant data breach involving several popular applications with embedded AI components. The incident exposed the personal information of over 500 million users. The database containing user details such as real names, usernames, gender, location and phone numbers was exposed and offered for sale~\cite{li2023data}. Since LLMs are trained on datasets and user-generated content from social media platforms like Facebook, such breaches raise significant concerns about privacy and data security in AI training processes, particularly problematic if the leaked data includes personal identifiers or proprietary details~\cite{kibriya2024privacy}.

\subsection{Regulatory and Policy Frameworks}

The ethical and privacy concerns associated with LLMs have prompted various regulatory responses worldwide. The European Union's proposed Artificial Intelligence Act aims to establish a comprehensive legal framework to ensure the ethical use of AI technologies, focusing on transparency, accountability, and human oversight. This legislation addresses the need for AI systems to disclose AI-generated content and provide explanations for AI-driven decisions, particularly in high-risk applications such as healthcare and law enforcement~\cite{madiega2021artificial}. In the United States, the Federal Trade Commission (FTC) has issued guidelines emphasizing fairness, transparency, and accountability in AI systems~\cite{jillson2021aiming}. 

Additionally, the National Institute of Standards and Technology (NIST) is developing a framework for AI risk management to provide organizations with practical tools for managing AI-related risks~\cite{ai2024artificial}.
China's Interim Measures for the Management of Generative AI Services, introduced in 2024, require clear labeling of AI-generated content and enforce strict data collection and usage guidelines to protect user privacy~\cite{migliorini2024china}. These regulatory efforts aim to balance innovation with ethical considerations, ensuring the responsible development and deployment of Generative AI technologies. Furthermore, industry-specific guidelines such as those by IEEE provide a framework for ethical AI design, highlighting principles like transparency, accountability, and data privacy~\cite{ieee2017ieee}.

Tech companies have also established their own ethical guidelines. For instance, Google's AI principles emphasize fairness, interpretability, privacy, and safety, aiming to prevent misuse of AI technologies~\cite{pichai2018ai}. OpenAI has similar guidelines focusing on long-term safety, robustness, and compliance with ethical standards~\cite{openai_safety_standards}.

Despite these regulatory advancements, challenges remain. Overly descriptive regulations might slow down innovation and make compliance burdensome for companies, particularly smaller enterprises that need to use LLMs. Additionally, the global nature of AI development necessitates international cooperation to integrate standards and ensure effective enforcement across jurisdictions. The varying regulatory landscapes across different countries can lead to inconsistencies and gaps in ethical standards and protections~\cite{floridi2018ai4people,buiten2019towards}. \textcolor{black}{The timeline of the development of regulatory and policy frameworks is shown in Table~\ref{tab:guidelines_timeline}.}

\textcolor{blue}{\begin{table}[!t]
  \centering
  \caption{Timeline of AI-Ethics Guidelines and Frameworks}
  \label{tab:guidelines_timeline}
  \begin{tabular}{|p{0.75\linewidth}|c|}
    \hline
    \textbf{Framework / Guideline}                                                    & \textbf{Year} \\ \hline
    AI Principles (Google)                                                             & 2018          \\ \hline
    OpenAI Charter (OpenAI)                                                            & 2018          \\ \hline
    Ethics Guidelines for Trustworthy AI (EU)                  & 2019          \\ \hline
    Ethically Aligned Design (IEEE)                                                 & 2019          \\ \hline
    FTC Guidelines               & 2021          \\ \hline
    Artificial Intelligence Act (EU)                                                   & 2021          \\ \hline
    AI Risk Management Framework (NIST)                                                & 2023          \\ \hline
    Interim Measures for Generative AI Services (China)                                & 2024          \\ \hline
  \end{tabular}
\end{table}}

\subsection{Industry and governmental guidelines selected}~\label{subsec:Frameworks}

As mentioned in the previous section, there have been multiple frameworks and guidelines produced to aid in identifying and addressing AI ethics and governance issues. We selected four guidelines/frameworks due to their availability, authority, and significant impact on shaping the ethical use of AI technologies.

\textbf{The IEEE Ethically Aligned Design Document} aims to establish ethical principles for Autonomous and Intelligent Systems (A/IS) that advance human beneficence, i.e., prioritize benefits to humanity and the environment, and mitigate risks associated with these technologies. It emphasizes that prioritizing human well-being must not conflict with environmental sustainability~\cite{ieee_ethically_aligned_design}. 

\textbf{National Institute of Standards and Technology (US)'s Artificial Intelligence Risk Management Framework}  outlines characteristics necessary for building trustworthy AI systems, recognizing both the transformative potential and unique risks of AI. The document emphasizes the importance of balancing various trustworthiness attributes to minimize potential harms~\cite{ai2023artificial}. 

\textbf{Microsoft Responsible AI Standard} aims to operationalize ethical principles across Microsoft’s AI development and deployment processes by providing concrete, actionable guidelines~\cite{microsoft_responsible_ai_standard}. The goals are organized into six key areas, each designed to support responsible and trustworthy AI.

\textbf{European Union (EU)'s AI Act} introduces a comprehensive regulatory framework to manage the development and deployment of AI across the EU, using a risk-based approach to tailor requirements for various AI applications~\cite{madiega2021artificial}.

\subsection{Existing Mitigation Strategies for Ethical Concerns}
Several mitigation strategies have been proposed and implemented to address the ethical concerns associated with LLMs. One key approach is the development of bias detection and mitigation techniques. Researchers have created tools such as the Large Language Model Bias Index (LLMBI) to quantify and address biases in LLM outputs~\cite{oketunji2023large}. These tools use advanced NLP techniques to detect and correct biases related to race, gender, and other sensitive attributes.

Another strategy involves enhancing transparency and interpretability through techniques like SHapley Additive exPlanations (SHAP) and Local Interpretable Model-agnostic Explanations (LIME), which provide insights into how models make decisions and allow users to understand and trust AI outputs~\cite{ribeiro2016should}. Privacy-preserving techniques such as differential privacy and federated learning are also being employed to protect individual data while training LLMs. Differential privacy introduces noise to the data to prevent the identification of specific individuals, while federated learning allows models to be trained across multiple decentralized devices without sharing raw data, thus enhancing privacy~\cite{mcmahan2017communication}.

Collaboration between AI developers, policymakers, and other stakeholders is crucial in creating ethical guidelines and regulatory frameworks that are both effective and adaptable. Industry-wide initiatives, such as the Partnership on AI, bring together diverse perspectives to address ethical challenges and promote best practices in AI development and deployment~\cite{partnership_on_ai_2021}. However, these strategies also have limitations. Bias detection tools can only address known biases and may miss subtle or emerging issues, potentially leading to ongoing ethical challenges~\cite{oketunji2023large}. Transparency techniques like SHAP and LIME can help interpret decisions but do not eliminate the fundamental complexity of LLMs, and their effectiveness relies on the correct application and understanding of these methods by users~\cite{ribeiro2016should}. Privacy techniques like differential privacy and federated learning require careful implementation to balance privacy with model performance, and there is limited empirical evaluation of their effectiveness in large-scale deployments~\cite{mcmahan2017communication}. Moreover, the current mitigation strategies often lack comprehensive evaluation, and their long-term impacts on AI ethics and performance remain uncertain~\cite{partnership_on_ai_2021}.

Overall, while these mitigation strategies are essential for addressing ethical concerns, they must be continuously refined and adapted to keep pace with the rapid advancements in AI technology. Collaborative efforts and ongoing research are vital to realising the benefits of Generative AI while minimizing potential harms.

\subsection{Related Systematic Reviews}~\label{subsec:relatedWork}

 The ethical implications of AI systems have garnered increasing attention in recent years, resulting in several systematic reviews and mapping studies. Li et al. (2023) conducted a systematic review focusing on the ethical concerns and related strategies for AI in healthcare, identifying 45 relevant studies. This review highlighted the need for transparency, privacy, and accountability in AI design and implementation, but it was primarily centered on healthcare applications, leaving a gap in understanding how these ethical concerns translate across other domains~\cite{li2022ethics}.
Atlam et al. (2024)  mapped AI ethics research over the past seven years and identified 127 primary studies. Their findings revealed a concentration of research on fairness, transparency, and accountability while pointing out a lack of empirical evidence supporting AI ethics principles~\cite{atlam2024slm}. However, the study was more of a high-level categorization, lacking in-depth analysis of specific methodologies used to address these ethical concerns.  A systematic literature review by the International Journal of Data Science and Analytics (2023) identified 66 papers focusing on developing objective metrics to assess the ethical compliance of AI systems~\cite{palumbo2024objective}. While this review underscored the need for standardized, quantifiable metrics to evaluate AI ethics, it was limited by its exclusion of frameworks that require human intervention, which might be necessary for a comprehensive understanding of ethical AI~\cite{palumbo2024objective}. Although these studies have provided valuable insights into the current development of AI ethics, their limitations highlight gaps in the literature. There is a need for a more comprehensive analysis that covers various domains of AI application when the focus is on healthcare, high-level categorizations and the exclusion of human-centred frameworks. This systematic mapping study aims to fill these gaps by providing a broader, more detailed exploration of AI ethics across diverse sectors, identifying the strategies used to address ethical concerns, and providing a foundation for the development of comprehensive, cross-domain ethical guidelines and frameworks.

\section{Methodology}~\label{sec:methodology}

We conducted a systematic mapping study (SMS) on the rising ethical concerns in using GenAI, particularly LLMs. We conducted our study on the scientific literature, existing frameworks, and guidelines, e.g., industry guidelines and governmental guidelines for AI ethics. Given the rapidly evolving nature of the field, we recognised that a comprehensive understanding of the ethical concerns surrounding LLMs cannot be derived solely from scientific literature. Therefore, we expanded our study to include industry frameworks, governmental guidelines, and other authoritative sources, ensuring a more holistic and up-to-date perspective on the ethical challenges in this dynamic area.
This SMS has been carried out in accordance with the guidelines for systematic mapping studies as outlined by Peterson et al.~\cite{petersen2015guidelines}. 

\begin{figure*}[h]
    \centering
    \includegraphics[width=\linewidth]{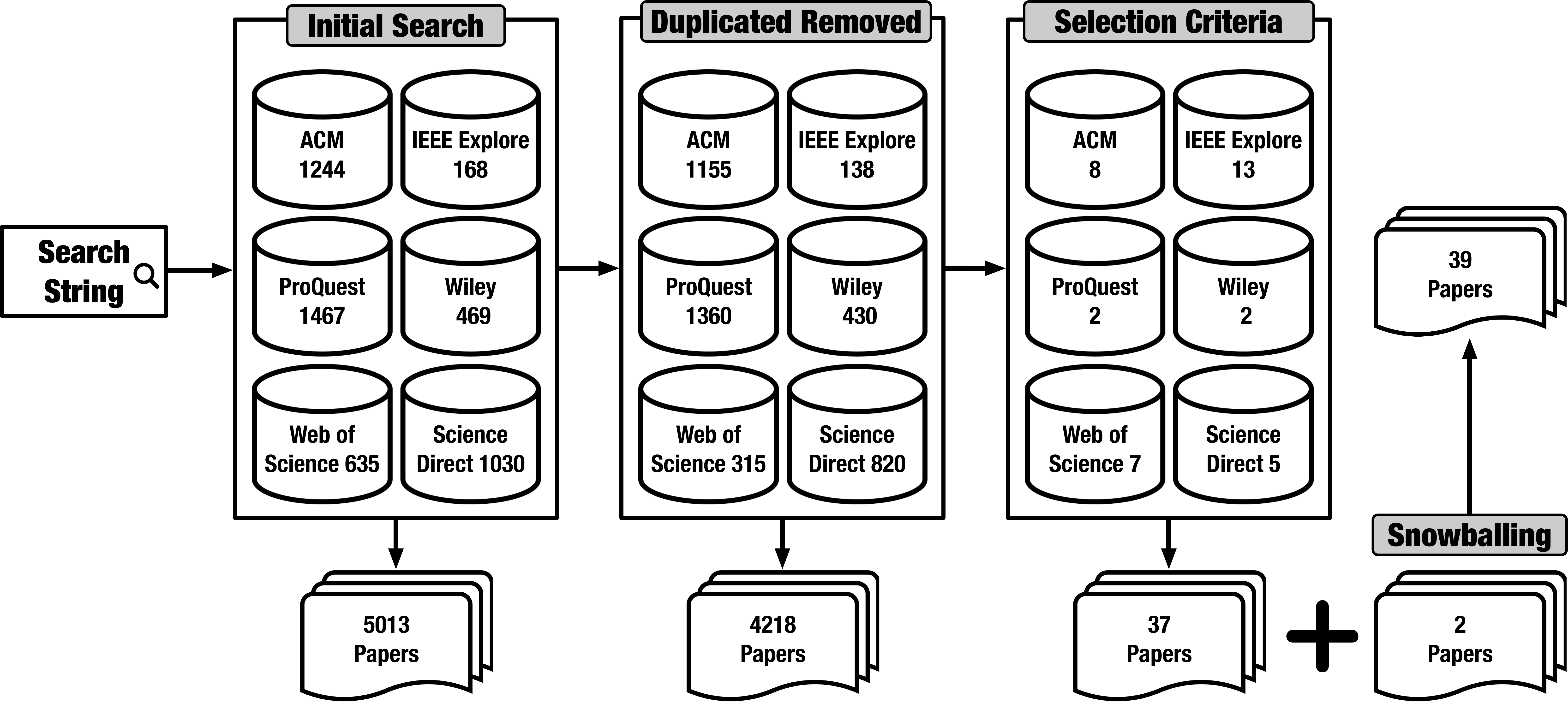}
    \caption{Study Search Process}
    \label{fig:search_process}
\end{figure*}

Our mapping study was conducted over three stages, i.e., planning, conducting, and reporting the review. The planning phase involved writing a protocol, formulating research questions (RQs) and reviewing our protocol for alignment to the main aim of our study. The protocol included a plan for a search strategy. 
During the conducting stage, the first author developed search strings that the other authors reviewed. A librarian from our university was also consulted at this stage to ensure alignment with the best practices for conducting systematic reviews. The search strings went through an iterative process, and some that were not specific or only marginally relevant were removed. The first author then led our search process (illustrated in Figure~\ref{fig:search_process}) with the search strings in various
databases and selected ACM Digital Library, IEEE Xplore, Proquest, Wiley, Web of Science, and Science Direct. The first author searched all selected databases and removed duplicates using Endnote in the initial paper screening.
We identified relevant keywords and search strings to explore existing empirical evidence exhaustively. We note that based on the guidelines in this stage, we used the search process to identify the relevant scientific literature. We identified the industry and governmental guidelines based on the references in the scientific literature and snowballing. \textcolor{black}{Following Wohlin's guidelines~\cite{wohlin2014guidelines}, we applied forward snowballing to the 37 studies remaining. Two researchers independently screened the citing articles against our predefined inclusion and exclusion criteria and resulted in 2 additional eligible studies.} In total, our search process led to 39 scientific studies and six industry and governmental guidelines. 

\subsection{Research Questions}
We formulated three high-level research questions (RQs): \\ 
RQ1 - What are the ethical dimensions defined in the use of generative AI across various fields?\\ 
RQ2 - What strategies are used or proposed to address the ethical concerns of using generative AI across various fields? \\
RQ3 - What are the challenges when implementing the strategies? \\
We use these RQs to identify studies related to the ethical dimensions of Generative AI usage and the corresponding strategies to address these dimensions. In our study, an ethical dimension refers to specific ethical considerations that arise from using GenAI or LLMs~\cite{sivasubramaniam2021assisting,poff2023academic,anderson2021ai}. These dimensions encompass a range of ethical concerns, principles, and guidelines crucial for the responsible development and use of LLMs.

\subsection{Search Strategy}

\subsubsection{Search String Formulation}

We formulated a search string to search online databases using key and alternative terms related to four main areas: large language models, guidelines, development, and ethics. However, we acknowledge that these areas may sometimes overlap with related concepts. Thus, we included a comprehensive set of terms in the search string to capture all relevant literature, shown in~\ref{AppendixB}. The search string was tested on six online databases: IEEE Xplore, ACM Digital Library, ProQuest, Web of Science, Wiley Online Library, and Science Direct. These databases were chosen because they are comprehensive and widely recognised sources of high-quality research in computer science, engineering, and technology. \textcolor{black}{In addition to these databases, we conducted a parallel search on Arxiv, finding three preprint studies that met our inclusion criteria. We included these papers to capture the most recent and relevant studies.} \textcolor{black}{We included the Arxiv papers as the topic of Ethical AI is relatively new, and Arxiv provides us with studies and research that may not yet have gone through peer review, which allows us to explore emerging ideas, diverse perspectives, and early-stage research that might not yet be available in journals and conferences.}

The ACM Digital Library and IEEE Xplore are particularly relevant as they contain a vast collection of
research papers and conference proceedings focused on AI and machine learning. ProQuest, Wiley, Web of Science, and ScienceDirect were included for their extensive coverage of multidisciplinary studies, including regulatory guidelines and ethical considerations in technology. Additionally, we included some of these databases to capture grey literature, e.g., the industry and governmental guidelines. Given that AI ethics is a relatively new field, grey literature (not part of the main 39 papers) provides valuable insights and complements our understanding. 

The search string was refined over several iterations to maximise the relevance of the results. For instance, the first author would randomly select a sample of 8-10 papers, review them for relevance check, and further refine the search string. 

In constructing the search strategies for various databases, we employed a range of logical connectors to ensure a comprehensive and relevant retrieval of literature on large language models and related issues. Each database required specific adjustments to optimize the search strategy, using various connectors and operators tailored to the database.

In IEEE, Web of Science, Wiley Online Library, and ScienceDirect, the search strategies made use of the ``AND'' connector to combine key concepts, ensuring that search results included multiple relevant topics such as governance, ethics, transparency, or accountability in relation to large language models (LLMs). The ``OR'' \textcolor{black}{operator} was used to include alternate expressions of similar concepts (e.g.,``ethic'' OR ``moral'', ``development'' OR ``design''). Additionally, symbols like the asterisk (*) were used to capture word variations. For instance, ``Large Language Model*'' ensures that both ``Large Language Model'' and ``Large Language Models'' are retrieved. This strategy helps retrieve documents that mention different term forms, maximizing the search's comprehensiveness.
In ACM Digital Library, the search strategy used the ``AND" \textcolor{black}{operator} and ``OR" \textcolor{black}{operator} to combine multiple concepts and capture alternate terms. ``AND'' was used to ensure that search results contained all the specified concepts, such as ``large language model*'' AND ``ethics'' AND ``governance.'' This ensures that the retrieved documents address all the selected topics together. ``OR'' was used to include variations or synonyms for a concept, such as ``ethics'' OR ``moral'' OR ``ethical'', which broadens the search to include documents that may use different terms for the same idea.

In ProQuest, we used a more refined approach by employing the ``noft'' \textcolor{black}{operator}, which stands for ``not in full text''. This operator excludes terms that appear only in the body of the document and not in more critical fields such as the title, abstract, or subject headings. The use of ``noft'' is beneficial in ProQuest because this database often contains a large volume of full-text content, including dissertations, reports, and articles that may mention key terms incidentally without focusing on them in depth. By using ``noft'', we can filter out documents where a term like ``governance'' is only briefly mentioned in the text but not central to the research focus. For example, using ``noft(governance)'' ensures that we retrieve documents where ``governance'' is emphasized in key fields, like the title or abstract, indicating that the topic is a primary focus of the work rather than a tangential mention.
This increases the relevance and specificity of the search results, making sure that only documents dealing substantially with ``governance'' are included. Additionally, the ``AND'' \textcolor{black}{operator} and ``OR'' \textcolor{black}{operator} were also used in ProQuest to combine different thematic elements and provide alternate terms for key concepts, ensuring comprehensive coverage of topics while maintaining the precision of the search.

\textcolor{black}{To include grey literature, we performed targetted searches of major AI ethics bodies' websites (IEEE, NIST, Microsoft, European Commission). Because these documents are not indexed in academic databases, no Boolean search string was applied; instead, we navigated each site and used keyword filtering (e.g., ``AI ethics” ``responsible AI”) in the site search tools. We found four guidelines/frameworks that can be used in the paper, which are included in Section~\ref{subsec:Frameworks}.}

\begin{table*}[h]
\centering
\caption{Inclusion and Exclusion Criteria}\label{table:Iinclusion_ExclusionCriteria}
\resizebox{0.9\textwidth}{!}{
\begin{tabularx}{\textwidth}{p{3cm} p{11.5cm}}
\toprule
\textbf{Criteria ID} & \textbf{Criterion} \\
\midrule
\multicolumn{2}{l}{\textbf{Inclusion Criteria}} \\
\midrule
I01 & Papers discussing ethical concerns in the use of Generative AI \\
I02 & Full text of the article is available. \\
I03 & Peer-reviewed studies, sector-specific studies \textcolor{black}{and grey literature(Arxiv, Guidelines and Frameworks)}. \\
I04 & Papers written in English language. \\
\midrule
\multicolumn{2}{l}{\textbf{Exclusion Criteria}} \\
\midrule
E01 & Papers about GenAI that do not discuss ethical concerns\\
E02 & Papers about ethical concerns that are not in the field of GenAI \\
E03 & Papers that are less than four pages in length. \\
E04 & Conference or workshop papers if an extended journal version of the same paper exists. \\
E05 & Papers with inadequate information to extract relevant data. \\
E06 & Vision papers,  books (chapters), posters, discussions, opinions, keynotes, magazine articles, experience, and comparison papers. \\
\bottomrule
\end{tabularx}
}

\end{table*}

\subsubsection{Automated Search and Filtering}
 Our search was conducted on databases between April and May 2024. The final search string was applied to the selected database online search engines, and an initial pool of 5013 papers was extracted. The list of papers was downloaded and exported to Endnote. Despite the large number of database results obtained by our search string (5013 papers), many papers were irrelevant or duplicates. 
 
 We first removed all duplicates (remaining papers = 4218) and applied a set of inclusion and exclusion criteria to filter out pertinent studies as part of our SMS protocol, shown in Table~\ref{table:Iinclusion_ExclusionCriteria}). For this mapping study, we only considered  English-language studies that directly discuss the ethical dimensions or concerns in their studies. Given that the topic is relatively new, we included short papers of more than four pages to harness emerging ideas. However, we excluded irrelevant studies such as posters, keynotes, opinion papers, and magazine articles. The selection criteria were applied to all studies to identify the most relevant ones, with discussions among all authors during the study filtering process. This left us with 39 papers remaining. 

\textcolor{black}{\subsubsection{Data Analysis Process}}

\begin{figure*}[h]
    \centering
    \includegraphics[width=1\linewidth]{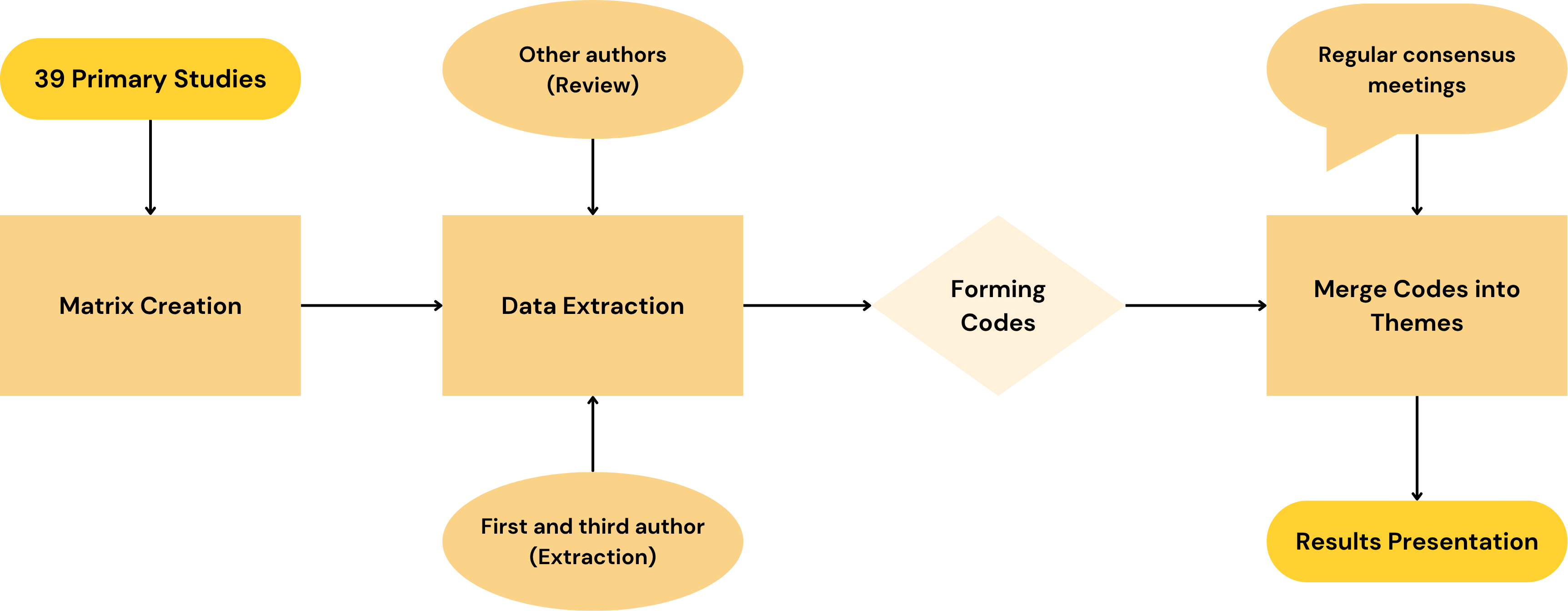}
    \caption{Data Analysis Process}
    \label{fig:data}
\end{figure*}

\textcolor{black}{Following the completion of our automated search and filtering, we conducted a structured analysis of the 39 selected primary studies. Our data analysis process is shown in Figure \ref{fig:data}. First, the first author created an article matrix to capture the studies' aims, methodologies, and key results; this matrix was then shared among all authors to establish an overview of the studies. The first and third authors then systematically identified and extracted data from each study that were relevant to our three RQs. The other authors independently reviewed these data extractions to verify the accuracy and comprehensiveness of the data. The first and third authors then organized the extracted data into preliminary categories aligned with our RQs. These categories were refined through an iterative process of condensing and open coding. We highlighted the relevant information for each RQ from the extracted data. We coded them according to the key information, and similar codes were then combined into themes. Throughout this phase, all authors met regularly to discuss coding decisions, reconcile discrepancies and reach consensus.}

\vspace{2\baselineskip}
\section{Results}~\label{sec:results}

\subsection{Selected Studies}
After filtering we selected 39 primary studies on ethical concerns surrounding the use of LLMs, published from 2020 to 2024, as illustrated in Figures 2 and 3. The bar chart in Figure 2 shows the year-wise distribution of the selected papers by their year of publication. Publications were sparse and relatively consistent from 2020 through 2022, with only minor fluctuations. However, 2023 saw a sharp increase, with 26 studies published, reflecting a substantial rise in research interest in ethical concerns in AI. This upward trend continued into 2024, though with a slight decrease, likely due to our research cutoff in June 2024.

Figure 3 displays the distribution of these studies by publication venue. Most studies, 56.4\%, were published in journals, followed by 33.3\% in conferences. Publications on Arxiv accounted for 7.7\%. This breakdown underscores that journals are the primary medium for disseminating research on AI ethics, while conferences also contribute significantly. We included the Arxiv papers as the topic of Ethical AI is relatively new, and Arxiv provides us with studies and research that may not yet have gone through peer review, which allows us to explore emerging ideas, diverse perspectives, and early-stage research that might not yet be available in journals and conferences.

\begin{figure*}[h]
    \centering
    \includegraphics[width=1\linewidth]{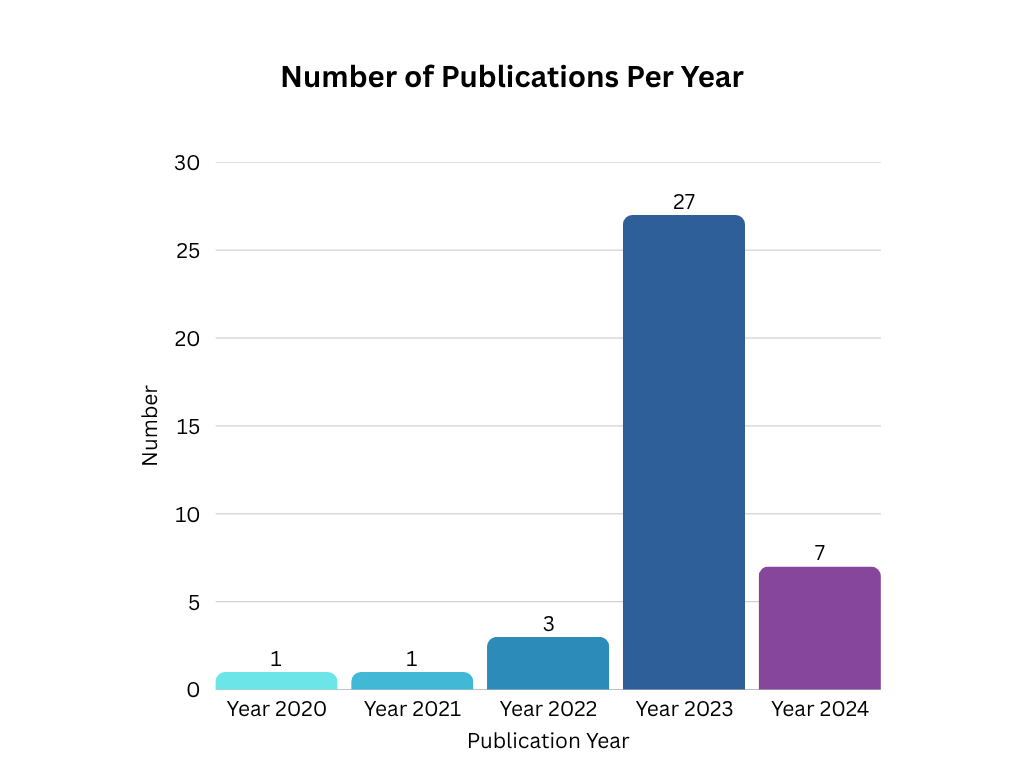}
    \caption{Study Distribution by Year}
    \label{fig:2}
\end{figure*}

\begin{figure*}[h]
    \centering
    \includegraphics[scale=0.5]{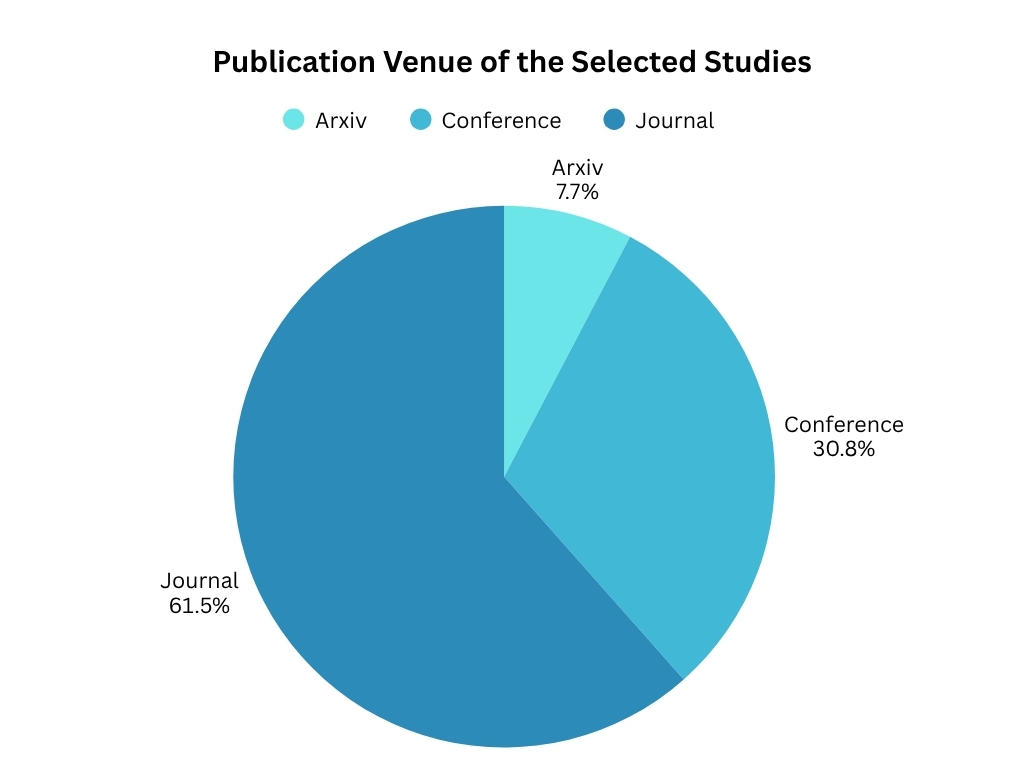}
    \caption{Study Distribution by publication}
    \label{fig:3}
\end{figure*}

%\subsection{Domains of Studies}

Our selected primary studies span multiple application domains. Each domain brings its own set of challenges and ethical considerations, reflecting the diverse applications and potential impacts of LLMs. This section categorizes the studies according to their respective domains mentioned in the primary studies, offering a detailed exploration of how ethical concerns manifest in different contexts. 

\begin{itemize}
    \item 
    \textit{General AI}: Papers categorized under ``General AI" address AI's broad concepts and applications without focusing on a specific domain. They discuss relevant ethical challenges across multiple sectors where AI is deployed.
\end{itemize}

\begin{itemize}
    \item 
    \textit{Healthcare}:  Healthcare is a domain that frequently engages with ethical concerns such as privacy, safety, and bias, especially with the increasing use of AI and LLMs for medical applications. The papers in this domain discuss how AI-driven systems can both enhance and complicate healthcare practices, with patient confidentiality, data security, and algorithmic fairness being recurring themes. This domain is highly sensitive due to the direct impact on human well-being and medical decision-making.
\end{itemize}

\begin{itemize}
    \item 
    \textit{Legal}: Papers focusing on the legal domain highlight the intersection of AI with law, exploring issues like accountability, transparency, and bias in legal AI systems. The integration of AI in legal contexts raises concerns about fairness, potential racial biases in predictive policing tools, and the transparency of AI-driven legal decisions.
\end{itemize}

\begin{itemize}
    \item 
    \textit{Education}:  The educational domain explores the ethical use of AI in learning environments, where concerns include privacy, fairness, and the transparency of AI systems in assessing student performance or providing educational content. 
\end{itemize}

\begin{itemize}
    \item 
    \textit{Public safety}: This domain involves the use of AI in public safety systems, where the ethical dimensions center on accountability, bias, and safety. AI systems used in policing, emergency response, and public surveillance must be scrutinized for fairness and transparency to avoid unintended harm to communities, especially marginalized groups. 
\end{itemize}

\begin{itemize}
    \item 
    \textit{Societal Impact}: Papers in this domain examine how LLMs interact with broader societal issues, such as their role in shaping public discourse, policy, and societal norms. Ethical concerns focus on the responsibility of those deploying LLMs to consider their social impact, including how they influence public opinion, access to information, and equality.
\end{itemize}

\begin{itemize}
    \item 
    \textit{Cybersecurity}: This domain highlights mainly the privacy concerns specifically related to systems that use AI components or LLMs. Papers here focus on issues such as the protection of personal data, the risks of data leakage, and how AI systems can be designed to respect user privacy, especially focusing on data-hungry models like LLMs.
\end{itemize}

\begin{itemize}
    \item 
    \textit{Economics}: Papers in this domain focus on the economic impact and ethical concerns surrounding AI and LLMs, particularly in relation to automation, job displacement, and the ethical use of AI in economic decision-making. The economic dimension of LLMs also raises issues of accessibility and fairness in how AI systems are deployed and who benefits from their use.
\end{itemize}

\textcolor{black}{The specific papers associated with each domain and ethical dimensions can be seen in Table \ref{table:domains}, organized into three columns. The first column ``Domain” groups the selected studies by domains. The second column ``Papers” lists each primary study in that domain. The third column ``Ethical Dimensions (n)” shows, for each domain, how many of its studies address each ethical dimension with the count. From the studies, it is worth noting that accountability is rarely addressed in the education and public safety domains, while both transparency and accountability are completely absent from the cybersecurity literature. Transparency also receives only minimal attention in the education, societal impact, legal, and public safety papers. It is critical that future research focuses on these underrepresented ethical dimensions within their respective sectors.}

\begin{table*}[!t]
  \centering
  \setlength{\extrarowheight}{5pt}
  \Large
  \caption{Domains mentioned in papers and their ethical‐dimension counts}
  \label{table:domains}
  \resizebox{\textwidth}{!}{%
    \begin{tabular}{|l|l|p{9cm}|}
      \hline
      \textbf{Domain} & \textbf{Papers} & \textbf{Ethical Dimensions (n)} \\
      \hline
      \textbf{Cybersecurity} 
        & [\textcolor{blue}{\ref{P1}}, \textcolor{blue}{\ref{P23}}, \textcolor{blue}{\ref{P27}}, \textcolor{blue}{\ref{P29}}, \textcolor{blue}{\ref{P31}}, \textcolor{blue}{\ref{P33}}] 
        & Safety (2), Privacy (6), Bias (1), Transparency (0), Accountability (0) \\ \hline
      \textbf{Education} 
        & [\textcolor{blue}{\ref{P14}}, \textcolor{blue}{\ref{P16}}, \textcolor{blue}{\ref{P26}}] 
        & Safety (2), Privacy (3), Bias (2), Transparency (2), Accountability (1) \\ \hline
      \textbf{General AI} 
        & [\textcolor{blue}{\ref{P2}}, \textcolor{blue}{\ref{P4}}, \textcolor{blue}{\ref{P5}},\textcolor{blue}{\ref{P15}}, \textcolor{blue}{\ref{P17}}, \textcolor{blue}{\ref{P20}}, \textcolor{blue}{\ref{P28}}, \textcolor{blue}{\ref{P30}}, \textcolor{blue}{\ref{P35}}, \textcolor{blue}{\ref{P37}}, \textcolor{blue}{\ref{P39}}] 
        & Safety (4), Privacy (7), Bias (8), Transparency (5), Accountability (4) \\ \hline
      \textbf{Healthcare} 
        & [\textcolor{blue}{\ref{P1}}, \textcolor{blue}{\ref{P9}}, \textcolor{blue}{\ref{P10}}, \textcolor{blue}{\ref{P11}}, \textcolor{blue}{\ref{P13}}, \textcolor{blue}{\ref{P21}}, \textcolor{blue}{\ref{P22}}, \textcolor{blue}{\ref{P24}}, \textcolor{blue}{\ref{P32}}, \textcolor{blue}{\ref{P34}}] 
        & Safety (7), Privacy (7), Bias (7), Transparency (7), Accountability (5) \\ \hline
      \textbf{Societal Impact} 
        & [\textcolor{blue}{\ref{P2}}, \textcolor{blue}{\ref{P25}}, \textcolor{blue}{\ref{P38}}] 
        & Safety (2), Privacy (3), Bias (2), Transparency (2), Accountability (1) \\ \hline
    \textbf{Legal} 
        & [\textcolor{blue}{\ref{P3}}, \textcolor{blue}{\ref{P6}}, \textcolor{blue}{\ref{P8}}] 
        & Safety (2), Privacy (1), Bias (3), Transparency (2), Accountability (1) \\ \hline
      \textbf{Public Safety} 
        & [\textcolor{blue}{\ref{P7}}, \textcolor{blue}{\ref{P18}}, \textcolor{blue}{\ref{P19}}, \textcolor{blue}{\ref{P27}}] 
        & Safety (4), Privacy (1), Bias (2), Transparency (1), Accountability (0) \\ \hline
    \end{tabular}%
  }
\end{table*}

\begin{table}[h]
    \centering
    \scriptsize
    \caption{Generative AI Technologies Used in Primary Studies}
    \begin{tabular}{|l|c|}
        \hline
        \textbf{LLM Models} & \textbf{Number Mentions} \\ \hline
        ChatGPT & 18 \\ \hline
        Conversational Agents & 6 \\ \hline
        DALL-E & 1 \\ \hline
        EduLLMs & 1 \\ \hline
        Gemini (Bard) & 2 \\ \hline
        GPT-2 & 1 \\ \hline
        General LLMs & 8 \\ \hline
        LLM-based Chatbots & 1 \\ \hline
        LLM Virtual Assistants & 1 \\ \hline
        RoBERTa & 1 \\ \hline
        Transformers & 1 \\ \hline
    \end{tabular}
    \label{tab:ai_technology_mentions}
\end{table}

We identified a diverse range of LLM models being used in the selected primary studies, shown in table \ref{tab:ai_technology_mentions}. A significant number of papers used versions of ChatGPT (\textcolor{blue}{\ref{P1}}, \textcolor{blue}{\ref{P7}}, \textcolor{blue}{\ref{P8}}, \textcolor{blue}{\ref{P11}}, \textcolor{blue}{\ref{P12}}, \textcolor{blue}{\ref{P14}}, \textcolor{blue}{\ref{P16}}, \textcolor{blue}{\ref{P17}}, \textcolor{blue}{\ref{P18}}, \textcolor{blue}{\ref{P19}}, \textcolor{blue}{\ref{P24}}, \textcolor{blue}{\ref{P25}}, \textcolor{blue}{\ref{P27}}, \textcolor{blue}{\ref{P28}}, \textcolor{blue}{\ref{P34}}, \textcolor{blue}{\ref{P35}}, \textcolor{blue}{\ref{P37}}, and \textcolor{blue}{\ref{P39}}), highlighting its versatile use across various domains such as healthcare, education, and general conversational interfaces.  
Conversational Agents (CA), another key area of focus, were discussed in multiple papers (\textcolor{blue}{\ref{P2}}, \textcolor{blue}{\ref{P5}}, \textcolor{blue}{\ref{P15}}, \textcolor{blue}{\ref{P29}}, \textcolor{blue}{\ref{P31}}, \textcolor{blue}{\ref{P36}}. These agents are noted for their applications in interactive and supportive roles, including customer service and educational tools.

General LLMs were featured in several papers (\textcolor{blue}{\ref{P4}}, \textcolor{blue}{\ref{P6}}, \textcolor{blue}{\ref{P20}}, \textcolor{blue}{\ref{P21}}, \textcolor{blue}{\ref{P30}}, \textcolor{blue}{\ref{P32}}, \textcolor{blue}{\ref{P33}}, \textcolor{blue}{\ref{P38}}). These papers emphasized the broad capabilities of LLMs in understanding and generating human-like text, highlighting their potential across various sectors. Additionally, LLM-based chatbots and virtual assistants were discussed in papers such as \textcolor{blue}{\ref{P9}}, \textcolor{blue}{\ref{P10}}, and \textcolor{blue}{\ref{P13}}, indicating their growing relevance in enhancing user interaction and automating responses.

Models such as GPT-2 (\textcolor{blue}{\ref{P3}}) and transformers (\textcolor{blue}{\ref{P22}}) were also explored, with RoBERTa being mentioned alongside ChatGPT in \textcolor{blue}{\ref{P23}} for the use of fine-tuning LLMs. Educational Large Language Models (EduLLMs) were specifically addressed in \textcolor{blue}{\ref{P26}}, showcasing their application in creating and facilitating educational content. Moreover, newer AI models like Gemini (Bard) and DALL-E were mentioned in the context of their capabilities and potential applications in papers \textcolor{blue}{\ref{P37}} and \textcolor{blue}{\ref{P39}}.

\subsection{RQ1 Results}

We wanted to identify the various ethical issues related to AI that have been discussed across the selected papers. The purpose was to explore both the breadth and depth of these issues and how they align with key ethical dimensions.

In defining the key ethical dimensions for our study, we focused on safety, accountability, bias, transparency, and privacy. These dimensions are central to the international frameworks and guidelines, discussed in Section~\ref{subsec:Frameworks}. \textcolor{black}{We selected these five dimensions because they are the most frequently emphasized across the four major frameworks in Section~\ref{subsec:Frameworks}, they also emerged as the most recurrent themes during our coding of the 39 primary studies. Other values such as fairness, autonomy and consent did appear in some of the studies, and they were included in our core categories. For example, fairness issues are coded under Bias, autonomy issues are related to Accountability, and consent-related issues are part of Privacy. Moreover, the EU’s seven Trustworthy AI requirements, such as “human agency and oversight” and “non-discrimination and fairness” are embedded in our dimensions. For instance, the EU’s “human agency” is captured by our Transparency (ensuring users understand AI decisions) and Accountability (clarifying who is responsible), while “non-discrimination” aligns with our analysis of Bias (algorithmic fairness). We specifically concentrated on these five dimensions as current LLMs need practical fixes like bias detection tools and privacy-preserving training methods, adding more standalone dimensions would risk diluting solutions for urgent issues.}

\begin{figure}[h]
    \centering    \includegraphics[width=1.0\textwidth]{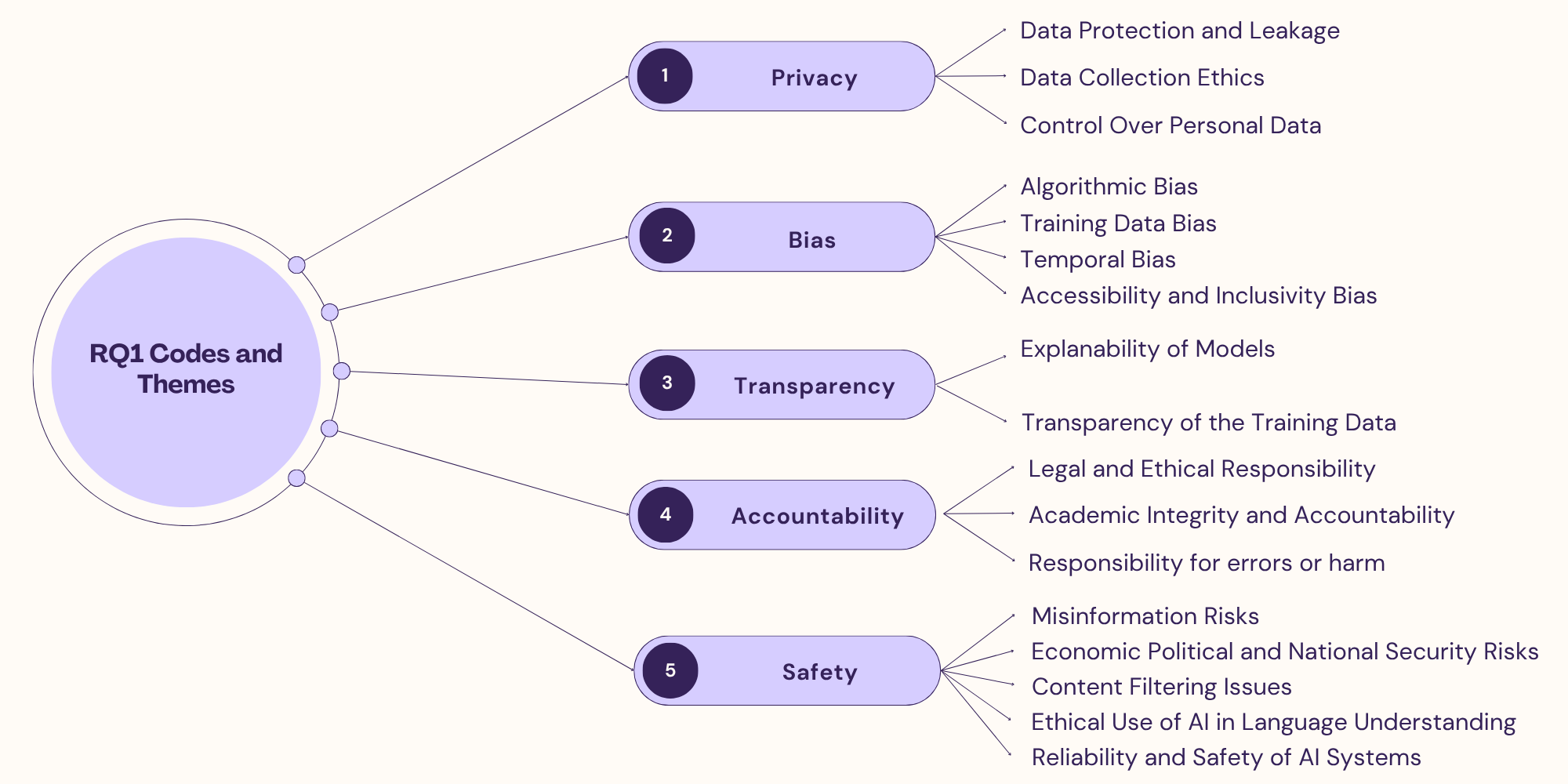}
    \caption{RQ1 codes and themes mapping}
    \label{fig:RQ1 Mapping}
\end{figure}

\textbf{Safety} is a prominent requirement in the EU AI Act, where high-risk AI systems must undergo thorough safety assessments to protect users and ensure system integrity in critical areas like healthcare and transportation. IEEE’s Ethically Aligned Design also emphasizes safety through the principle of beneficence, which prioritizes human well-being in A/IS design  \cite{madiega2021artificial}. Our focus on safety incorporates the need for operational robustness and user protection across different contexts, recognizing it as a foundational element of trustworthy AI.

\textbf{Accountability} also emerged as a crucial dimension, highlighted in both the NIST AI Risk Management Framework and Microsoft Responsible AI Standard, which call for strong accountability measures to hold developers and operators responsible for AI system outcomes. NIST’s framework identifies accountability as key to effective risk management, requiring organizations to account for both foreseeable and unforeseen impacts \cite{ai2023artificial}. By selecting accountability, we underscore the importance of defining ethical boundaries and responsibilities within AI operations.

\textbf{Bias} is widely acknowledged as an ethical concern due to the potential for AI systems to propagate discrimination and inequity. This dimension is explicitly covered in frameworks like IEEE’s Ethically Aligned Design and the NIST AI RMF, which both stress the importance of reducing algorithmic bias to ensure fair, equitable outcomes \cite{ieee_ethically_aligned_design,ai2023artificial}. Bias mitigation is especially pertinent in high-stakes areas, such as hiring and law enforcement, where biased AI outcomes can have a direct and significant impact on society.

\textbf{Transparency} is a universally emphasized dimension, with all four frameworks underscoring the need for AI systems to be explainable, interpretable, and auditable. The EU AI Act enforces transparency requirements for high-risk applications, demanding that developers provide clear explanations of AI decision-making processes to foster trust \cite{madiega2021artificial}. Similarly, NIST lists transparency as fundamental to AI trustworthiness, enabling stakeholders to understand and evaluate AI operations. This dimension is central to building user trust and ensuring regulatory oversight, both necessary for responsible AI deployment \cite{ai2023artificial}.

\textbf{Privacy} remains a critical concern in AI ethics, particularly with the influence of GDPR embedded within the EU AI Act, which enforces stringent data protection for AI systems \cite{madiega2021artificial}. NIST’s framework also highlights privacy-enhanced AI, advocating for systems that protect personal data and uphold users’ informational rights \cite{ai2023artificial}. By focusing on privacy, we address data misuse risks and safeguard users’ rights to control their information.

We used these five overarching ethical dimensions to categorize studies addressing key ethical concerns related to LLM usage. We also considered the perspectives of the authors in each paper, incorporating their opinions on which ethical dimension specific issues should belong to. This ensured that our categorization was informed not only by the frameworks and guidelines but also by the authors' interpretations. The analysis involved carefully reviewing each paper and coding the mentioned ethical issues, which were then grouped into themes that fit within these categories. In total, we identified 130 different ethical issues across the studies, which we categorized into 17 distinct themes. Figure \ref{fig:RQ1 Mapping}
 shows the key ethical dimensions to theme mappings. Table \ref{ethics} shows the primary studies addressing each dimension.

 \subsubsection{Safety} 
The theme \emph{``Ethical Use of AI in Language Understanding"} highlights the ethical challenges related to how AI systems, particularly virtual assistants (VAs), process and respond to language, including inappropriate or harmful speech, this theme includes \textcolor{blue}{\ref{P10}}. In \textcolor{blue}{\ref{P10}}, a key concern is raised about the need to train VAs with inappropriate language for them to recognize and understand it. This raises ethical questions about whether the VA should be able to recognise insults or offensive language and, if so, how it should respond. 

\begin{table}[h]
\centering
\footnotesize
\begin{longtable}{p{3cm} p{6cm} p{4cm}}
    \caption{Ethical Dimensions Identified from papers}\\
    \toprule
    \textbf{Ethical \newline Dimensions} & \textbf{Definition} & \textbf{Paper ID} \\
    \midrule
    \endfirsthead

    \caption[]{Ethical Dimensions Identified from papers (continued)} \\
    \toprule
    \textbf{Ethical \newline Dimensions} & \textbf{Definition} & \textbf{Paper ID} \\
    \midrule
    \endhead

    \midrule
    \endfoot

    \bottomrule
    \endlastfoot

    \textbf{Safety} & Generative AI systems need to be ensured to operate safely and securely, minimizing risks and issues & \textcolor{blue}{\ref{P2}}, \textcolor{blue}{\ref{P6}}, \textcolor{blue}{\ref{P7}}, \textcolor{blue}{\ref{P8}}, \textcolor{blue}{\ref{P9}}, \textcolor{blue}{\ref{P10}}, \textcolor{blue}{\ref{P12}}, \textcolor{blue}{\ref{P13}}, \textcolor{blue}{\ref{P14}}, \textcolor{blue}{\ref{P18}}, \textcolor{blue}{\ref{P19}}, \textcolor{blue}{\ref{P21}}, \textcolor{blue}{\ref{P22}}, \textcolor{blue}{\ref{P25}}, \textcolor{blue}{\ref{P27}}, \textcolor{blue}{\ref{P28}}, \textcolor{blue}{\ref{P30}}, \textcolor{blue}{\ref{P32}}, \textcolor{blue}{\ref{P34}}, \textcolor{blue}{\ref{P38}} \\
    \midrule
    \textbf{Privacy} & The information and privacy of individuals need to be protected when using generative AI systems & \textcolor{blue}{\ref{P1}}, \textcolor{blue}{\ref{P2}}, \textcolor{blue}{\ref{P5}}, \textcolor{blue}{\ref{P7}}, \textcolor{blue}{\ref{P8}}, \textcolor{blue}{\ref{P9}}, \textcolor{blue}{\ref{P10}}, \textcolor{blue}{\ref{P11}}, \textcolor{blue}{\ref{P12}}, \textcolor{blue}{\ref{P14}}, \textcolor{blue}{\ref{P16}}, \textcolor{blue}{\ref{P17}}, \textcolor{blue}{\ref{P22}}, \textcolor{blue}{\ref{P23}}, \textcolor{blue}{\ref{P24}}, \textcolor{blue}{\ref{P25}}, \textcolor{blue}{\ref{P26}}, \textcolor{blue}{\ref{P27}}, \textcolor{blue}{\ref{P28}}, \textcolor{blue}{\ref{P29}}, \textcolor{blue}{\ref{P31}}, \textcolor{blue}{\ref{P33}}, \textcolor{blue}{\ref{P34}}, \textcolor{blue}{\ref{P37}}, \textcolor{blue}{\ref{P38}}, \textcolor{blue}{\ref{P39}} \\
    \midrule
    \textbf{Transparency} & The operations and decisions of generative AI systems are open and understandable to stakeholders & \textcolor{blue}{\ref{P2}}, \textcolor{blue}{\ref{P5}}, \textcolor{blue}{\ref{P6}}, \textcolor{blue}{\ref{P8}}, \textcolor{blue}{\ref{P10}}, \textcolor{blue}{\ref{P11}}, \textcolor{blue}{\ref{P12}}, \textcolor{blue}{\ref{P13}}, \textcolor{blue}{\ref{P15}}, \textcolor{blue}{\ref{P16}}, \textcolor{blue}{\ref{P19}}, \textcolor{blue}{\ref{P21}}, \textcolor{blue}{\ref{P22}}, \textcolor{blue}{\ref{P24}}, \textcolor{blue}{\ref{P34}}, \textcolor{blue}{\ref{P38}}, \textcolor{blue}{\ref{P39}} \\
    \midrule
    \textbf{Bias} & The biases of the system need to be addressed as they may affect the fairness and equity of the AI system & \textcolor{blue}{\ref{P1}}, \textcolor{blue}{\ref{P3}}, \textcolor{blue}{\ref{P4}}, \textcolor{blue}{\ref{P5}}, \textcolor{blue}{\ref{P6}}, \textcolor{blue}{\ref{P8}}, \textcolor{blue}{\ref{P10}}, \textcolor{blue}{\ref{P11}}, \textcolor{blue}{\ref{P12}}, \textcolor{blue}{\ref{P14}}, \textcolor{blue}{\ref{P15}}, \textcolor{blue}{\ref{P16}}, \textcolor{blue}{\ref{P17}}, \textcolor{blue}{\ref{P18}}, \textcolor{blue}{\ref{P19}}, \textcolor{blue}{\ref{P20}},  \textcolor{blue}{\ref{P22}}, \textcolor{blue}{\ref{P24}}, \textcolor{blue}{\ref{P25}}, \textcolor{blue}{\ref{P28}}, \textcolor{blue}{\ref{P30}}, \textcolor{blue}{\ref{P34}}, \textcolor{blue}{\ref{P35}}, \textcolor{blue}{\ref{P36}}, \textcolor{blue}{\ref{P38}} \\
    \midrule
    \textbf{Accountability} & AI systems need to be responsible and their actions can be traced and justified & \textcolor{blue}{\ref{P1}}, \textcolor{blue}{\ref{P5}}, \textcolor{blue}{\ref{P11}}, \textcolor{blue}{\ref{P16}}, \textcolor{blue}{\ref{P17}}, \textcolor{blue}{\ref{P21}}, \textcolor{blue}{\ref{P24}}, \textcolor{blue}{\ref{P34}}, \textcolor{blue}{\ref{P37}}, \textcolor{blue}{\ref{P38}}\\
\end{longtable}
\label{ethics}
\end{table}

The theme \emph{``Misinformation Risk Issues"} identifies the ethical concerns related to AI systems generating and disseminating false or misleading information, which can have serious consequences. This theme was covered in \textcolor{blue}{\ref{P2}}, \textcolor{blue}{\ref{P6}}, \textcolor{blue}{\ref{P8}}, \textcolor{blue}{\ref{P10}}, \textcolor{blue}{\ref{P12}}, \textcolor{blue}{\ref{P13}}, \textcolor{blue}{\ref{P14}}, \textcolor{blue}{\ref{P28}}, \textcolor{blue}{\ref{P30}}, \textcolor{blue}{\ref{P32}}, \textcolor{blue}{\ref{P38}}. In \textcolor{blue}{\ref{P12}}, a significant risk is highlighted with ChatGPT potentially producing sophisticated but hallucinatory fake information, especially in financial outputs, which could be hard to detect and lead to substantial financial losses. In \textcolor{blue}{\ref{P30}}, another concern is raised about LLMs generating misinformation, resulting in less informed users and eroding public trust in shared information.

The theme \emph{``Economic, Political, and National Security Risks"} focuses on the dangers posed by AI systems in sensitive areas, particularly regarding the misuse of generative AI like ChatGPT. this theme is covered in \textcolor{blue}{\ref{P18}}. In \textcolor{blue}{\ref{P18}}, political security risks are noted. AI could inadvertently access and assimilate sensitive information, such as national secrets, trade secrets, or personal data, through user inputs, potentially leading to information leakage. Additionally, military security is threatened by the potential misuse of AI to generate attack codes targeting critical national infrastructure or military facilities. Economic security is also at risk, as tampering with training data could create false financial data and market predictions, undermining economic integrity. 

The theme \emph{``Content Filtering Issues"} addresses the ethical concern of LLMs, processing both accurate and harmful content due to inadequate filtering mechanisms. This theme covers \textcolor{blue}{\ref{P13}}. In \textcolor{blue}{\ref{P13}}, it is noted that LLMs can ingest and incorporate harmful content along with accurate information, which can result in the AI producing outputs that may be inappropriate or damaging.

The theme \emph{``Reliability and Safety of AI Systems"} focuses on the ethical concerns surrounding the accuracy and dependability of AI, particularly in critical situations; this theme is covered in \textcolor{blue}{\ref{P7}}, \textcolor{blue}{\ref{P9}}, \textcolor{blue}{\ref{P13}}, \textcolor{blue}{\ref{P14}}, \textcolor{blue}{\ref{P19}}, \textcolor{blue}{\ref{P21}}, \textcolor{blue}{\ref{P25}}, \textcolor{blue}{\ref{P27}}, \textcolor{blue}{\ref{P28}}, \textcolor{blue}{\ref{P30}}, \textcolor{blue}{\ref{P34}}. In \textcolor{blue}{\ref{P19}}, a significant issue is the provision of oversimplified and erroneous advice on safety matters, with the advice often lacking traceability due to missing cited sources, making fact‐checking difficult. This raises concerns about AI’s reliability in delivering accurate and trustworthy information in safety‐critical contexts. In \textcolor{blue}{\ref{P21}}, the issue of trust is discussed, especially regarding the reliability of LLMs in situations outside their training conditions. This phenomenon, known as an ``out‐of‐distribution shift,'' can lead to performance failures.

\subsubsection{Privacy} 

The theme \emph{``Control over Personal Data"} emphasizes the ethical importance of protecting individuals' privacy and ensuring they have control over how their personal information is used, particularly in AI-driven systems. As AI increasingly relies on sensitive data, especially in fields like healthcare, safeguarding this data becomes crucial to prevent misuse and ensure privacy compliance, this theme is covered in \textcolor{blue}{\ref{P1}}, \textcolor{blue}{\ref{P14}} and \textcolor{blue}{\ref{P31}}. In \textcolor{blue}{\ref{P1}}, one key concern is the need for robust data protection and transparency in AI's decision-making, especially in healthcare, to avoid bias and ensure fairness. In \textcolor{blue}{\ref{P14}}, an example highlights the potential for adding privacy settings to give users control over how their data is collected and shared, reinforcing the need for individual autonomy over personal information.

The theme \emph{``Data Protection and Leakage"} addresses the ethical concerns related to the privacy and security of data in AI systems, particularly regarding the risk of unauthorized access or data breaches, this theme is covered in \textcolor{blue}{\ref{P1}}, \textcolor{blue}{\ref{P2}}, \textcolor{blue}{\ref{P5}}, \textcolor{blue}{\ref{P7}}, \textcolor{blue}{\ref{P8}}, \textcolor{blue}{\ref{P9}}, \textcolor{blue}{\ref{P10}}, \textcolor{blue}{\ref{P11}}, \textcolor{blue}{\ref{P12}}, \textcolor{blue}{\ref{P14}}, \textcolor{blue}{\ref{P16}}, \textcolor{blue}{\ref{P17}}, \textcolor{blue}{\ref{P22}}, \textcolor{blue}{\ref{P23}}, \textcolor{blue}{\ref{P24}}, \textcolor{blue}{\ref{P25}}, \textcolor{blue}{\ref{P26}}, \textcolor{blue}{\ref{P27}}, \textcolor{blue}{\ref{P28}}, \textcolor{blue}{\ref{P29}}, \textcolor{blue}{\ref{P30}}, \textcolor{blue}{\ref{P33}}, \textcolor{blue}{\ref{P34}}, \textcolor{blue}{\ref{P37}}, \textcolor{blue}{\ref{P38}}, \textcolor{blue}{\ref{P39}}. In \textcolor{blue}{\ref{P23}}, the issue is highlighted in the context of AI researchers fine-tuning pre-trained models with private data for various tasks. These models are vulnerable to privacy attacks due to their tendency to memorize training data, known as the "memorization issue," posing risks to sensitive information. In \textcolor{blue}{\ref{P26}}, concerns arise around smart education systems, which collect and analyze vast amounts of student data to provide personalized learning. 

The theme \emph{``Data Collection Ethics"} focuses on the ethical issues surrounding how data is gathered for training LLMs. This theme is covered in \textcolor{blue}{\ref{P14}} and \textcolor{blue}{\ref{P27}}. In \textcolor{blue}{\ref{P27}}, concerns are raised about scraping information from internet forums and other sources without proper consent or oversight, which raises questions about the ethicality of the data collection process. In \textcolor{blue}{\ref{P14}}, another issue is the lack of disclosing the underlying sources used in LLM training, as they are not shared with the public.

\subsubsection{Bias}

The theme \emph{``Algorithmic Bias"} highlights the ethical concerns regarding how AI models can perpetuate social stereotypes and discrimination, particularly when trained on biased data. This theme is covered in \textcolor{blue}{\ref{P4}}, \textcolor{blue}{\ref{P5}}, \textcolor{blue}{\ref{P6}}, \textcolor{blue}{\ref{P12}}, \textcolor{blue}{\ref{P22}}, \textcolor{blue}{\ref{P24}}, \textcolor{blue}{\ref{P25}}, \textcolor{blue}{\ref{P28}}, \textcolor{blue}{\ref{P30}}, \textcolor{blue}{\ref{P34}}, \textcolor{blue}{\ref{P35}}, \textcolor{blue}{\ref{P38}}. In \textcolor{blue}{\ref{P25}}, an example is provided where Amazon used AI in human resources, and it was found that women were consistently rated lower than men. This bias arose because the AI analyzed resumes from the past decade, a period where most resumes came from men, indicating that the data was neither large nor diverse enough to mitigate bias. In \textcolor{blue}{\ref{P30}}, a similar concern is raised about how LLMs can perpetuate social stereotypes and introduce representational and allocational harms, leading to discrimination.

The theme \emph{``Training Data Bias''} addresses the ethical concerns that arise when AI models are trained on datasets that either overrepresent or underrepresent certain demographic groups. This theme is covered in \textcolor{blue}{\ref{P1}}, \textcolor{blue}{\ref{P3}}, \textcolor{blue}{\ref{P4}}, \textcolor{blue}{\ref{P8}}, \textcolor{blue}{\ref{P10}}, \textcolor{blue}{\ref{P11}}, \textcolor{blue}{\ref{P16}}, \textcolor{blue}{\ref{P17}}, \textcolor{blue}{\ref{P28}}, \textcolor{blue}{\ref{P35}}. In \textcolor{blue}{\ref{P35}}, demographic bias is highlighted as a major issue, where this imbalance in the data can cause models to exhibit biased behaviors toward specific genders, races, ethnicities, or other social groups. In \textcolor{blue}{\ref{P28}}, the concern is further emphasized, showing how biased data can perpetuate social stereotypes and lead to unfair discrimination. When models are trained on skewed representations of certain groups, it can result in unjust outcomes, reinforcing existing inequalities.

The theme \emph{``Temporal Bias''} focuses on the ethical concerns arising from the time-based limitations of training data used in AI models. This theme is associated with \textcolor{blue}{\ref{P35}}. In \textcolor{blue}{\ref{P35}}, it is noted that these models are often trained on data from specific time periods or have temporal cutoffs, which can lead to bias when reporting on current events, trends, or opinions. This limitation may result in the model providing outdated or inaccurate information when addressing contemporary topics and an incomplete understanding of historical contexts due to the lack of temporally diverse data. This can affect the relevance and accuracy of AI-generated content in time-sensitive situations.

The theme \emph{``Accessibility and Inclusivity Bias''} highlights the ethical concerns related to how AI systems may overlook the needs of individuals with disabilities or those from diverse linguistic backgrounds, this theme is covered in \textcolor{blue}{\ref{P14}}, \textcolor{blue}{\ref{P15}}, \textcolor{blue}{\ref{P18}}, \textcolor{blue}{\ref{P19}}, \textcolor{blue}{\ref{P20}}, \textcolor{blue}{\ref{P35}}. In \textcolor{blue}{\ref{P14}}, it is highlighted that accessibility features, such as screen readers, alternative text for images, or video captions, are often inadequate or absent in AI systems. Additionally, limitations in language translation capabilities may exclude non-native speakers or those with diverse linguistic needs.

\subsubsection{Accountability}

The theme \emph{``Legal and Ethical Responsibility"} focuses on the challenges surrounding accountability and moral responsibility when using AI in decision-making processes, especially in critical contexts, this theme is covered in \textcolor{blue}{\ref{P5}}, \textcolor{blue}{\ref{P21}}, \textcolor{blue}{\ref{P34}}, \textcolor{blue}{\ref{P37}}, \textcolor{blue}{\ref{P38}}. In \textcolor{blue}{\ref{P5}}, it is noted that accountability is a pivotal element in AI governance, particularly when delegating tasks such as predictions or decisions to AI systems. However, the definition of accountability in AI remains ambiguous and should be clarified based on factors such as the subject, scope, and context of its application. In \textcolor{blue}{\ref{P21}}, the issue of moral responsibility arises in high-stakes decisions, such as medical care, where traditionally, a human is held accountable. When an AI algorithm is involved, it becomes less clear who is responsible for the outcome, especially in adverse decisions. Establishing clear guidelines for when a human professional may ``overrule" the AI is essential to maintaining accountability.

The theme \emph{``Academic Integrity and Accountability"} highlights the ethical concerns surrounding using AI, particularly ChatGPT, in academic settings, this theme is covered in \textcolor{blue}{\ref{P11}}, \textcolor{blue}{\ref{P16}}, \textcolor{blue}{\ref{P17}}. In \textcolor{blue}{\ref{P16}}, the broader academic community has raised concerns about students using ChatGPT to plagiarize assignments, research papers, and other academic work, which undermines academic integrity. In \textcolor{blue}{\ref{P17}}, questions arise regarding the proper use of ChatGPT’s responses for academic purposes, such as whether AI-generated content should be credited and how accountability is managed if it is misused. The uncertainty extends to legal implications, mainly if AI-generated material is used maliciously or in ways that breach academic or ethical standards.

The theme \emph{``Responsibility for Errors or Harm"} involves the ethical concerns related to AI-driven decisions and the risks associated with responsibility. This theme is covered in \textcolor{blue}{\ref{P1}} and \textcolor{blue}{\ref{P24}}. In \textcolor{blue}{\ref{P1}}, it is highlighted that the availability of AI and automated decision aids can lead to a human tendency to rely too heavily on these technologies, often minimizing cognitive effort. In healthcare, this over-reliance on AI-driven decisions by clinicians can lead to misleading conclusions, potentially endangering patient safety and well-being.

\subsubsection{Transparency}

The theme of \emph{``Explainability of Models''} addresses the transparency challenges posed by the complexity of LLMs. In \textcolor{blue}{\ref{P38}}, the ``black box'' nature of LLMs is emphasized, highlighting the difficulty in understanding their decision-making processes due to the vast number of parameters they contain—often in the millions or billions. This theme is covered in \textcolor{blue}{\ref{P2}}, \textcolor{blue}{\ref{P5}}, \textcolor{blue}{\ref{P8}}, \textcolor{blue}{\ref{P10}}, \textcolor{blue}{\ref{P11}}, \textcolor{blue}{\ref{P12}}, \textcolor{blue}{\ref{P13}}, \textcolor{blue}{\ref{P15}}, \textcolor{blue}{\ref{P16}}, \textcolor{blue}{\ref{P21}}, \textcolor{blue}{\ref{P22}}, \textcolor{blue}{\ref{P24}}, \textcolor{blue}{\ref{P34}}, \textcolor{blue}{\ref{P38}}, and \textcolor{blue}{\ref{P39}}. This complexity makes it nearly impossible to explain how these models generate their outputs fully. In \textcolor{blue}{\ref{P39}}, the mystery surrounding the emergent capabilities of LLMs is further discussed as researchers remain uncertain about what drives these behaviors. 

The theme \emph{``Transparency of the Training Data''} focuses on the ethical issues arising from the lack of clarity regarding the data used to train AI models. This theme is covered in \textcolor{blue}{\ref{P6}} and \textcolor{blue}{\ref{P19}}. In \textcolor{blue}{\ref{P6}}, it is noted that the training data is often treated as a trade secret, and while efforts are underway to reverse engineer what data was used, companies neither confirm nor deny the accuracy of these guesses. This opacity raises concerns about the integrity and representativeness of the training data. In \textcolor{blue}{\ref{P19}}, the lack of transparency becomes particularly problematic when ChatGPT provides answers or recommendations, as users are left without a clear understanding of the sources behind its advice. Unlike human experts who can cite specific references, ChatGPT operates on a probabilistic model, making it difficult to trace or explain the origins of its responses.

\Needspace{10\baselineskip}
\begin{fancybox}[RQ1 Key Takeaways]
The RQ1 findings reveal critical ethical concerns across safety, transparency, accountability, privacy, and bias. Safety issues, such as misinformation risks and reliability, highlight the need for secure, trustworthy AI systems in high-stakes areas. Transparency challenges, including model explainability and unclear training data, reflect ongoing difficulties in understanding LLM decisions. Accountability concerns raise questions about defining responsibility in AI-driven decisions. Privacy issues emphasize user control over personal data in AI applications, while bias concerns rooted in algorithmic and training data show how LLMs risk reinforcing societal inequalities.
\end{fancybox}

\subsection{RQ2 Results}

\begin{table}[h]
  \scriptsize
  \setlength{\tabcolsep}{6pt}                  % reduce horizontal padding
  \caption{Mapping of RQ2 Themes to Ethical Dimensions}
  \label{tab:rq2matrix}
  \begin{tabular}{|p{0.4\linewidth}|c|c|c|c|c|}
    \hline
    \textbf{Theme}                                       & \textbf{Accountability} & \textbf{Transparency} & \textbf{Bias} & \textbf{Privacy} & \textbf{Safety} \\[.8ex]\hline
    Ethical Frameworks \& Interdisciplinary Collaboration                   & \checkmark    & \checkmark      &              &               &              \\[.8ex]\hline
    Accountability \& Continuous Oversight in AI systems               & \checkmark    &      &    \checkmark           &               &              \\[.8ex]\hline
    Bias Mitigation \& Fairness in AI systems                          &               &                 & \checkmark   &               &              \\[.8ex]\hline
    User Empowerment \& Transparency in AI Interactions               &               & \checkmark      &              & \checkmark    &              \\[.8ex]\hline
    Enhancing Trust \& Interpretability in AI systems            &       \checkmark         &                 &              &               & \checkmark   \\[.8ex]\hline
  \end{tabular}
\end{table}

To address the second research question, ``What strategies are used to address the ethical dimensions?", we conducted a thorough review of the studies. We extracted the mitigation strategies and recommendations presented in each paper. A mitigation strategy refers to a specific approach or action proposed or implemented to directly address or reduce a particular risk, issue, or challenge. In contrast, a recommendation is a suggestion or guidance proposed by the authors for future actions, often highlighting potential solutions or best practices that do not require immediate implementation. These strategies were categorized accordingly, and we further explored their evaluation status. \textcolor{black}{To ensure consistent classification, we developed a rubric: Not evaluated: the study simply proposes a mitigation strategy without any empirical test or user feedback, Fully evaluated: the strategy has been subjected to quantitative or qualitative assessment, such as user studies, case trials, or benchmark experiments, with clear outcome measures. We also categorized the studies that did not do a full evaluation on the strategies, but only partially evaluated, meaning the mitigation strategy had undergone some empirical investigation, but lacked full quantitative/qualitative or clear outcome measure as not evaluated. The authors independently applied this rubric to all strategies identified.} This process allowed us to systematically identify and classify the approaches proposed to address the ethical dimensions of AI use. The mitigation strategies and recommendations were \textcolor{black}{fully} evaluated only in 5/39 (\textcolor{black}{12.8\%}) studies. After identifying the evaluation status, we conducted a thematic analysis and categorized various mitigation strategies or recommendations, reflecting the proposed or practical solutions presented. Table \ref{tab:rq2matrix} shows the mapping of RQ2 themes to ethical dimensions.

\begin{itemize}
    \item 
    \textit\emph{Ethical Frameworks and Interdisciplinary Collaboration}: Establishing ethical guidelines for LLMs requires input from diverse disciplines to address the complex ethical challenges they present, this theme is covered in \textcolor{blue}{\ref{P1}}, \textcolor{blue}{\ref{P2}}, \textcolor{blue}{\ref{P4}}, \textcolor{blue}{\ref{P6}}, \textcolor{blue}{\ref{P16}}, \textcolor{blue}{\ref{P18}}, \textcolor{blue}{\ref{P24}}, \textcolor{blue}{\ref{P25}}, \textcolor{blue}{\ref{P30}}, \textcolor{blue}{\ref{P35}}, \textcolor{blue}{\ref{P37}}, \textcolor{blue}{\ref{P38}}. In \textcolor{blue}{\ref{P1}}, the recommendation emphasizes the need to establish ethical guidelines and regulatory frameworks for AI in healthcare, advocating for interdisciplinary collaboration to ensure that ethical standards are comprehensive and adaptable to advancements in technology. This collaboration is essential to create frameworks that are both effective and applicable across various domains. In \textcolor{blue}{\ref{P2}}, a Machine Ethics approach is proposed, suggesting that ethical standards and reasoning should be directly embedded within AI systems. This approach would enable AI to make ethically informed decisions autonomously, integrating ethical principles into the core of AI functionality. In \textcolor{blue}{\ref{P6}}, there is a focus on the legal dimensions of AI, suggesting that the training phase of AI may be covered under fair use; however, clearer guidelines are needed to inform system users. These examples illustrate the necessity of establishing ethical frameworks that are continually refined through interdisciplinary cooperation, ensuring that LLMs operate ethically and align with societal expectations.

\end{itemize}

\begin{itemize}
    \item 
    \textit{Bias Mitigation and Fairness in AI systems}: Addressing biases and promoting fairness in LLMs requires a multifaceted approach, with strategies implemented at various stages of the AI development process, this theme is covered in \textcolor{blue}{\ref{P1}}, \textcolor{blue}{\ref{P2}}, \textcolor{blue}{\ref{P3}}, \textcolor{blue}{\ref{P4}}, \textcolor{blue}{\ref{P6}}, \textcolor{blue}{\ref{P8}}, \textcolor{blue}{\ref{P16}}, \textcolor{blue}{\ref{P19}}, \textcolor{blue}{\ref{P25}}, \textcolor{blue}{\ref{P28}}, \textcolor{blue}{\ref{P35}}. In \textcolor{blue}{\ref{P3}}, a mitigation strategy is employed using a fill-in-the-blank method in a GPT-2 model to ensure that context is carefully considered during predictions, focusing on binary racial decisions between ``White" and ``Black". This method involves examining racial bias by masking racial references within the context, prompting the model to assign probabilities and observe potential biases. In \textcolor{blue}{\ref{P4}}, a combination of techniques is used to handle biases systematically: pre-processing methods transform input data to reduce biases before training, while in-processing techniques modify learning algorithms to eliminate discrimination during training. Additionally, post-processing methods are applied to adjust the model's output after training, particularly when retraining is not an option, treating the model as a black box. These strategies showcase a comprehensive effort to mitigate biases at different levels of the AI development pipeline, aiming to create fairer and more equitable AI systems.

\end{itemize}

\begin{itemize}
    \item 
    \textit{Enhancing Trust and Interpretability in AI Systems}: Building trust in LLMs relies heavily on making AI systems transparent and understandable to users, especially in critical fields like healthcare, this theme includes \textcolor{blue}{\ref{P1}}, \textcolor{blue}{\ref{P7}}, \textcolor{blue}{\ref{P9}}, \textcolor{blue}{\ref{P15}}, \textcolor{blue}{\ref{P20}}, \textcolor{blue}{\ref{P24}}, \textcolor{blue}{\ref{P30}}. In \textcolor{blue}{\ref{P1}}, a recommendation is made to focus on improving the interpretability of AI algorithms so that healthcare professionals can clearly understand and trust the decisions generated by these systems. This involves making the decision-making processes of AI transparent, allowing professionals to verify and rely on AI outputs confidently. In \textcolor{blue}{\ref{P7}}, a mitigation strategy involves using leading questions to guide ChatGPT when it fails to generate valid responses, enabling users to refine the system's outputs until they are accurate iteratively. This interaction fosters trust by giving users greater control over the AI’s response quality. In \textcolor{blue}{\ref{P9}}, the HCMPI method is recommended to reduce data dimensions, focusing on extracting only relevant K-dimensional information for healthcare chatbot systems. This reduction makes the AI’s reasoning clearer and more interpretable, allowing users to follow the underlying logic without getting overwhelmed by excessive data. These strategies aim to enhance AI systems by making them more interactive and interpretable.

\end{itemize}

\begin{itemize}
    \item 
    \textit{User Empowerment and Transparency in AI Interactions}: Empowering users and ensuring transparency in AI systems are critical for fostering ethical interactions and trust, this theme is covered in \textcolor{blue}{\ref{P1}}, P2 \textcolor{blue}{\ref{P2}}, \textcolor{blue}{\ref{P5}}, \textcolor{blue}{\ref{P8}}, \textcolor{blue}{\ref{P10}}, \textcolor{blue}{\ref{P11}}, \textcolor{blue}{\ref{P12}}, \textcolor{blue}{\ref{P13}}, \textcolor{blue}{\ref{P16}}, \textcolor{blue}{\ref{P17}}, \textcolor{blue}{\ref{P24}}, \textcolor{blue}{\ref{P28}}, \textcolor{blue}{\ref{P29}}, \textcolor{blue}{\ref{P30}}, \textcolor{blue}{\ref{P31}}, \textcolor{blue}{\ref{P33}}. In \textcolor{blue}{\ref{P2}}, a recommendation is made to emphasize critical reflection throughout conversational AI's design and development phases. This includes giving users more control over their interactions with AI agents and being transparent about the AI’s non-human nature and limitations. Such transparency allows users to understand the AI's capabilities and limitations, enabling more informed decision-making. In \textcolor{blue}{\ref{P8}}, a mitigation strategy involves techniques like logit output verification and proactive detection of hallucinations. These strategies are paired with participatory design, where users actively shape AI systems, ensuring that the AI’s responses remain accurate and meaningful. In \textcolor{blue}{\ref{P30}}, further mitigation strategies include functionality audits to assess whether LLM applications meet their intended goals and impact audits to evaluate how AI affects users, specific groups, and the broader environment. These strategies prioritize user involvement and clarity, fostering a transparent and user-centered AI ecosystem.

\end{itemize}

\begin{itemize}
    \item 
    \textit{Accountability and Continuous Oversight in AI Systems}: Ensuring that AI systems are accountable and continuously monitored is essential to maintain ethical standards and protect users, this theme is covered in \textcolor{blue}{\ref{P5}}, \textcolor{blue}{\ref{P7}}, \textcolor{blue}{\ref{P9}}, \textcolor{blue}{\ref{P10}}, \textcolor{blue}{\ref{P11}}, \textcolor{blue}{\ref{P12}}, \textcolor{blue}{\ref{P13}}, \textcolor{blue}{\ref{P20}}, \textcolor{blue}{\ref{P24}}, \textcolor{blue}{\ref{P25}}, \textcolor{blue}{\ref{P30}}, \textcolor{blue}{\ref{P31}}, \textcolor{blue}{\ref{P32}}, \textcolor{blue}{\ref{P35}}. In \textcolor{blue}{\ref{P24}}, a recommendation is made for AI tools like ChatGPT to be evaluated by regulators, specifically in healthcare settings, to ensure safety, efficacy, and reliable performance. This highlights the importance of oversight from regulatory bodies to ensure that AI applications adhere to established standards. In \textcolor{blue}{\ref{P9}}, a mitigation strategy involves the Healthcare Chatbot-based Zero Knowledge Proof (HCZKP) method, which enables the use of data without making it visible. This approach reduces the need for extensive data collection, safeguarding privacy, and ensuring ethical data handling. Additionally, the strategy recommends adopting data minimization principles by collecting only necessary essential data and decentralizing patient data during feedback training. These strategies emphasize the need for ongoing oversight and accountability in how AI systems manage and utilize data, particularly in sensitive environments.

\end{itemize}

Although many recommendations remain conceptual, several aforementioned mitigation strategies have been empirically assessed in the studies and have demonstrated promise in real-world scenarios. The pre-processing, in-processing and post-processing techniques have shown to reduce demographic disparities (\textcolor{blue}{\ref{P4}}), the HCMPI method and the HCZKP protocol have been deployed to limit sensitive patient data exposure (\textcolor{blue}{\ref{P9}}), the ethical information seeking framework that balanced cognitive load was pilot tested in an online learning environment, showing feasibility (\textcolor{blue}{\ref{P14}}). These concrete success cases underscore the value of the strategies for domain-specific deployment and also highlight the priorities for future empirical validation.

\Needspace{8\baselineskip}
\begin{fancybox}[RQ2 key Takeaways]
The findings from RQ2 identify various mitigation strategies across the five ethical dimensions of bias, transparency, accountability, privacy, and safety. Strategies include bias mitigation techniques to promote fairness, enhancing interpretability to build transparency and user trust, and fostering accountability through regulatory oversight. Privacy is safeguarded through data minimization practices, while safety is reinforced by interdisciplinary frameworks that adapt ethical standards to evolving AI technologies.
\end{fancybox}

\subsection{RQ3 Results}

\begin{table}[h]
  \scriptsize
  \centering
  \caption{Mapping of RQ3 Themes to Ethical Dimensions}
  \label{tab:rq3matrix}
  \setlength{\tabcolsep}{2pt}  
  \resizebox{\columnwidth}{!}{%
    \begin{tabular}{|p{0.35\columnwidth}|c|c|c|c|c|}
      \hline
      \textbf{Theme}                   & \textbf{Accountability} & \textbf{Transparency} & \textbf{Bias} & \textbf{Privacy} & \textbf{Safety} \\[.6ex]\hline
      Data-Related Challenges          &               &               &              & \checkmark    & \checkmark    \\[.6ex]\hline
      Technical Challenges             &               &               & \checkmark   & \checkmark    & \checkmark    \\[.6ex]\hline
      Legality Challenges              &               &              & \checkmark   &    & \checkmark    \\[.6ex]\hline
      Ethical Standard Challenges      &          \checkmark     &    &    &     &        \checkmark       \\[.6ex]\hline
      Individual Use Challenges        & \checkmark    &     &   \checkmark            &     &               \\[.6ex]\hline
      Ethical Dilemma Challenges       &     & \checkmark    &    \checkmark          &               &               \\[.6ex]\hline
    \end{tabular}%
  }
\end{table}

While numerous mitigation strategies have been proposed to address the ethical concerns surrounding the use of LLMs, many studies indicate that significant challenges persist even after these strategies are implemented. Authors in several papers have explicitly acknowledged that the mitigation efforts, while promising, often fall short due to various technical, legal, and governance barriers. These challenges highlight the complexity of achieving effective ethical governance in LLM applications. To better understand these issues, we have categorized the identified challenges into key themes, each reflecting the unique difficulties encountered during the practical implementation of these mitigation strategies.

\begin{itemize}
    \item 
    \textit{Technical Challenges}: Technical challenges in implementing mitigation strategies for LLMs often revolve around the complexity of integrating advanced techniques within diverse and sensitive contexts, this theme is covered in \textcolor{blue}{\ref{P1}}, \textcolor{blue}{\ref{P2}}, \textcolor{blue}{\ref{P4}}, \textcolor{blue}{\ref{P7}}, \textcolor{blue}{\ref{P10}}, \textcolor{blue}{\ref{P11}}, \textcolor{blue}{\ref{P12}}, \textcolor{blue}{\ref{P14}}, \textcolor{blue}{\ref{P20}}, \textcolor{blue}{\ref{P21}}, \textcolor{blue}{\ref{P22}}, \textcolor{blue}{\ref{P25}}, \textcolor{blue}{\ref{P28}}, \textcolor{blue}{\ref{P29}}, \textcolor{blue}{\ref{P30}}, \textcolor{blue}{\ref{P32}}, \textcolor{blue}{\ref{P33}}. In \textcolor{blue}{\ref{P1}}, the use of Natural Language Processing (NLP) on Electronic Health Records (EHR) data highlights multiple challenges, such as accurately identifying clinical entities, ensuring privacy protection, handling spelling errors, and managing the de-identification of sensitive information. These challenges are compounded by the lack of labeled data, the difficulty of detecting negations, and the complexities of deciphering numerous medical abbreviations. In \textcolor{blue}{\ref{P2}}, Gonen and Goldberg (2019) critique the current debiasing methods applied to word vectors. While these techniques produce high scores on self-defined metrics, they often result in only superficially hiding biases rather than genuinely eliminating them. \textcolor{blue}{\ref{P25}} emphasizes that Red Teaming, a method used to identify security vulnerabilities, must be a continuous process and not a one-time solution. It requires a persistent commitment to security throughout the development and operation of AI systems, with careful consideration of both legal and ethical implications. These examples demonstrate the multifaceted technical hurdles that complicate the practical application of ethical mitigation strategies in LLMs.

\end{itemize}

\begin{itemize}
    \item 
    \textit{Ethical Standard Challenges}: Implementing ethical standards in the context of LLMs is another challenging task, particularly due to the dynamic and diverse regulatory and cultural environments in which these technologies operate, this theme inclues \textcolor{blue}{\ref{P2}}, \textcolor{blue}{\ref{P8}}, \textcolor{blue}{\ref{P10}}, \textcolor{blue}{\ref{P30}}, \textcolor{blue}{\ref{P32}}, \textcolor{blue}{\ref{P34}}. In \textcolor{blue}{\ref{P10}}, the proposed recommendations are based on the analysis of the current regulatory landscape and anticipated regulations by the European Commission. However, as regulations continue to evolve, these recommendations may need to be revised to remain applicable and relevant. This highlights the importance of staying updated with the latest legal requirements in each jurisdiction to ensure ongoing compliance. \textcolor{blue}{\ref{P34}} underscores the significance of incorporating a user-centered approach in cybersecurity, recognizing that addressing human factors can enhance the effectiveness of these measures. Yet, as cybersecurity threats evolve, there is a need for continuous development of strategies to stay ahead of emerging challenges. In \textcolor{blue}{\ref{P8}}, it is noted that a one-size-fits-all approach to ethical standards is unlikely to be effective, as LLMs span diverse cultural and global contexts. The absence of standardized or universally regulated frameworks for LLMs adds to the complexity, requiring adaptable and context-sensitive solutions. These examples illustrate the challenges of creating consistent and effective ethical standards in a rapidly changing and diverse landscape.

\end{itemize}

\begin{itemize}
    \item 
    \textit{Ethical Dilemma Challenges}: Ethical dilemmas in the implementation of LLMs often arise from conflicting values and considerations, especially when navigating fairness, transparency, and privacy. In \textcolor{blue}{\ref{P3}}, the issue of learned associations within AI models highlights a significant ethical dilemma, this theme is covered in \textcolor{blue}{\ref{P3}}, \textcolor{blue}{\ref{P5}}, \textcolor{blue}{\ref{P8}}, \textcolor{blue}{\ref{P9}}, \textcolor{blue}{\ref{P10}}. For example, in \textcolor{blue}{\ref{P3}}, language models trained on data that frequently describe suspects and criminals as ``black males'' may reinforce stereotypes, raising concerns about legal equality. This brings up the ethical question of whether such associations, which exist in the training data, should be allowed in AI systems deployed for legal purposes, where fairness and equality are paramount. In \textcolor{blue}{\ref{P5}}, another ethical dilemma is identified in balancing transparency and protecting non-public information. While transparency is crucial for understanding an AI system’s behavior and decision-making process, companies often prioritize commercial secrecy to protect proprietary technology and business models. These conflicts illustrate the complexities of ethical decision-making in AI, where satisfying one ethical principle may compromise another, necessitating careful consideration and context-specific solutions.

\end{itemize}

\begin{itemize}
    \item 
    \textit{Individual Use Challenges}: Implementing mitigation strategies for LLMs encounters specific challenges related to how individuals use and interpret AI-generated content, this theme is covered in \textcolor{blue}{\ref{P13}}, \textcolor{blue}{\ref{P25}}, \textcolor{blue}{\ref{P27}}, \textcolor{blue}{\ref{P30}}. In \textcolor{blue}{\ref{P13}}, even with established guidelines and recommendations, reports have surfaced in the healthcare and wellness sectors where individuals consult LLMs like ChatGPT for health-related matters and take its advice without proper scrutiny. Despite mitigation efforts, the widespread sharing of such AI-generated content on social media demonstrates the difficulty in ensuring that users critically assess AI advice. In \textcolor{blue}{\ref{P25}}, engaging multiple stakeholders introduces further complications, even when measures are in place to involve diverse perspectives. Some stakeholders may use their involvement to advance personal agendas, reinforce existing biases, or misuse sensitive data, undermining transparency and ethical intentions. These challenges illustrate the complexities in managing individual behavior and ensuring responsible AI use, despite mitigation strategies aimed at fostering ethical and informed engagement with LLMs.

\end{itemize}

\begin{itemize}
    \item 
    \textit{Legality Challenges}: Legal challenges in implementing mitigation strategies for LLMs often involve navigating the complexities of intellectual property, censorship, and transparency, this theme is covered in \textcolor{blue}{\ref{P7}}, \textcolor{blue}{\ref{P8}}, \textcolor{blue}{\ref{P30}}. In \textcolor{blue}{\ref{P7}}, one significant concern is the potential leakage of business secrets and proprietary information when users interact with LLMs like ChatGPT. If proprietary code is inadvertently shared during AI interactions, it may become part of the chatbot's knowledge base, raising issues of copyright infringement and the preservation of business confidentiality. This poses a risk for organizations that depend on protecting sensitive information. In \textcolor{blue}{\ref{P8}}, censorship within LLMs, while intended to prevent harmful outputs, introduces legal dilemmas. There is often no clear or objective standard for determining what content is harmful, which can lead to the suppression of free speech or creative expression. Overly restrictive censorship may also hinder important debates, while a lack of transparency around censorship policies can create distrust in the AI system. These examples underscore the legal complexities of implementing effective mitigation strategies, where balancing ethical considerations with regulatory compliance is a persistent challenge.

\end{itemize}

\begin{itemize}
    \item 
    \textit{Data-related Challenges}: Data-related challenges are a critical factor in the implementation of mitigation strategies for LLMs, as they affect the quality, availability, and reliability of the datasets used in AI development, this theme is covered in \textcolor{blue}{\ref{P11}}, \textcolor{blue}{\ref{P12}}, \textcolor{blue}{\ref{P28}}, \textcolor{blue}{\ref{P32}}. In \textcolor{blue}{\ref{P28}}, concerns are raised about the future limitations of data collection and usage in machine learning. Research indicates that high-quality language data could be exhausted by 2026, with lower-quality data potentially running out by 2060. This forecast suggests that the limited availability of suitable datasets may constrain the future development and improvement of LLMs, affecting their ability to perform effectively and ethically. In \textcolor{blue}{\ref{P32}}, the quality of data is further questioned, as a substantial portion of source material comes from preprint servers that lack rigorous peer review. This reliance on unverified data can limit the generalizability and reliability of LLMs, particularly when data is drawn from diverse and variable contexts. These examples highlight the significant data challenges faced when attempting to implement mitigation strategies, where the quality, scope, and future availability of data play a crucial role in shaping the effectiveness of LLM interventions.

\end{itemize}

\begin{samepage}
\begin{fancybox}[RQ3 Key Takeaways]
    The RQ3 findings reveal significant challenges in applying mitigation strategies for LLMs, spanning technical, ethical, individual, legal, and data-related obstacles. Technical challenges arise from the complexity of integrating mitigation techniques in diverse and sensitive contexts, while maintaining ethical standards proves difficult across varied regulatory landscapes and cultural settings. Ethical dilemmas further complicate implementation, as strategies often conflict with core values, e.g., bias and transparency. Individual use challenges underscore the difficulty of ensuring users critically engage with AI-generated content, even when guidelines are in place. Legal barriers around intellectual property, censorship, and transparency introduce further complexity, and data-related issues, including data quality and availability, present constraints for sustainable LLM development.
\end{fancybox}
\end{samepage}

\section{Discussion}~\label{sec:discussion}

Our systematic mapping study reveals that the ethical concerns surrounding the deployment and usage of LLMs are multifaceted and have evolved significantly over recent years. This complexity is reflected in the increasing volume of research and the diverse range of mitigation strategies aimed at addressing these ethical dimensions. In this section, we analyze the prominent ethical issues identified, including safety, transparency, accountability, privacy, and bias. These dimensions, discussed across the selected studies, underscore a need for robust ethical frameworks and regulatory oversight to responsibly guide LLM deployment. Additionally, we discuss the challenges encountered in implementing these strategies, highlighting gaps that require further research and practical solutions to support ethical AI deployment.

\subsection{Contextual significance of ethical dimensions accross different domains}

In examining the RQ1 data, we noticed that the significance of each ethical dimension—safety, privacy, accountability, bias, and transparency—varies across domains. This suggests that the ethical concerns identified span a range of relevance depending on the application context.
For instance, safety is particularly emphasized in public safety applications within the education sector, where autonomous systems must prioritize safeguarding individuals, especially in student-focused scenarios~\cite{leslie2019understanding}. Conversely, safety appears less frequently in general AI applications where the primary concerns shift towards bias and transparency, suggesting that safety might hold less immediate significance in these broader AI use cases~\cite{floridi2022unified}.

Privacy emerges as a critical issue in healthcare, where data sensitivity is high, particularly for applications in AI privacy frameworks and general use of LLMs~\cite{jobin2019global}. This contrasts with economics-focused applications, where privacy remains essential but is balanced against other ethical considerations, such as transparency in auditing contexts. In healthcare, protecting user data aligns with legal mandates such as HIPAA, making privacy a non-negotiable dimension. In economic applications, transparency takes precedence due to regulatory oversight needs~\cite{dignum2019responsible}.
Bias shows a strong presence in legal and economic contexts, specifically regarding predictive models that could exacerbate racial or socioeconomic disparities. Bias is not as prominently discussed in more general AI applications, where transparency and interpretability may be prioritized over fairness, especially in domains where the risks associated with bias are not as pronounced~\cite{hagendorff2020ethics}.
Transparency is crucial across most domains but is particularly highlighted in the use of LLMs for public-facing applications, such as education and healthcare, where trustworthiness directly impacts user experience and compliance, e.g., in economic applications, transparency is essential in auditing AI outputs, ensuring that stakeholders can trust and verify model decisions~\cite{morley2020initial}.
Accountability, while universally relevant, is most emphasized in contexts with high legal or public stakes, such as healthcare and government AI applications, where the ability to trace and attribute decisions is essential for compliance and ethical standards~\cite{anawati2024artificial}. In more experimental applications, accountability is less prioritized, with a stronger focus instead on developing fundamental functionalities~\cite{smith2021clinical}.

\subsection{The need for Continuous Evaluation and Iterative Improvements}

The ethical concerns surrounding the use of LLMs, including bias, privacy, and transparency, are complex and continuously evolving. These issues are compounded by the dynamic nature of LLM deployments, which means that ethical strategies cannot remain static but must adapt to emergent challenges. Although frameworks such as the EU AI Act and the National Institute of Standards and Technology (NIST) emphasize the importance of responsive, adaptable frameworks, the real-world application of this principle remains limited. Most current ethical approaches primarily focus on initial implementation, with less attention to iterative evaluation, leaving gaps in sustained effectiveness. For instance, Nigam Shah et al. highlight the critical need for ongoing assessment in AI prediction algorithms to ensure long-term reliability and ethical alignment~\cite{shah2024need}. 

An additional concern is the lack of empirical evaluations of these strategies. Despite the recognized need for robust testing of bias mitigation, privacy safeguards, and transparency practices, relatively few systematic studies assess these methods in practical, high-stakes environments~\cite{palumbo2024objective}. This lack of empirical evidence means many strategies are based on theoretical or conceptual frameworks rather than validated, real-world outcomes. Consequently, organizations may find it difficult to anticipate how well these ethical measures will hold up over time or under evolving conditions.

The shifting landscape of AI ethics, driven by advances in data science and changing societal norms, also creates challenges for maintaining ethical standards. Strategies designed for earlier LLMs may not be effective for models trained on more diverse or sensitive datasets, where biases may arise unexpectedly, or privacy risks could be amplified. As AI continues to develop, the frameworks that govern its deployment must keep pace, ensuring that ethical practices evolve alongside technological progress~\cite{ortega2024applying}. The limitations of ``checkbox" ethics, as noted by Sara Kijewski et al., further highlight the importance of deeper, more adaptable standards that extend beyond initial implementation~\cite{kijewski2024rise}.

Frameworks like Microsoft's AI Standard and IEEE's guidelines advocate for continuous accountability, emphasising that ethical oversight should be an ongoing process rather than a one-time checkpoint. These frameworks aim to create adaptable, accountable structures that can evolve in response to new insights and societal changes~\cite{srikumar2022advancing}. However, implementing such adaptable strategies is challenging and requires a commitment to monitoring, reviewing, and refining ethical measures throughout the lifecycle of an AI system. The role of regular auditing, as highlighted by Li and Goel, is particularly essential in supporting accountability~\cite{li2024making}.

\begin{tcolorbox}[title=Recommendation, breakable]
    We recommend that mitigation approaches focus on addressing emerging biases, security vulnerabilities, and aligning with legal standards, particularly those related to evolving datasets. The use of ethical strategies, which are any planned combination of policies, processes and technical measures designed to mitigate a specific ethical concern should not remain static. Still, it should adapt over time as standards like EU AI Act and NIST are on multi-year review cycles. If ethical strategies remain static beyond review intervals, they risk becoming outdated due to technical advances and regulatory requirements. It is essential to prioritize accountability and support the development of flexible and evolving frameworks and guidelines.
\end{tcolorbox}

\subsection{Achieving Cross-Framework Consistency} 

Implementing ethical standards in AI is often complicated by the lack of harmonized guidelines across various frameworks and domains. For instance, the European Union's AI Act provides explicit regulatory requirements for privacy and bias mitigation, whereas frameworks from organizations like the NIST and IEEE offer guidelines that are often voluntary and context-specific~\cite{prem2023ethical}. This discrepancy can lead to inconsistencies, especially when LLMs are deployed across regions with differing regulatory landscapes.

The absence of standardized global ethical principles poses significant challenges for international organisations. Without a unified foundation, companies may struggle to ensure consistency in ethical AI practices, leading to potential ethical lapses and legal complications~\cite{khan2023ai}. This fragmentation is evident in the varying approaches to AI ethics across different jurisdictions, as highlighted by Hagendorff, who notes that the proliferation of AI ethics guidelines has led to a ``bewildering variety" of recommendations, often lacking coherence and practical applicability~\cite{hagendorff2020ethics}.

Moreover, the dynamic nature of AI technologies necessitates adaptable ethical frameworks that can evolve with technological advancements. However, the current lack of cross-framework consistency hampers the development of such adaptable guidelines. Vakkuri and Kemell emphasize the importance of implementing AI ethics in practice through empirical evaluation, yet they acknowledge the challenges posed by the absence of standardized ethical frameworks~\cite{vakkuri2019implementing}.

\begin{tcolorbox}[title=Recommendation]
    \textcolor{black}{We recommend establishing a core set of globally standardized ethical principles for the design, development, deployment and use of AI systems, providing a unified foundation on which different frameworks can build tailored guidance for specific domains.} This coordinated approach would support companies operating across jurisdictions by ensuring consistency in ethical AI practices while allowing flexibility for domain-specific requirements.
\end{tcolorbox}

\subsection{The lack of Scalability of Mitigations}

Implementing ethical mitigation strategies for LLMs presents significant scalability challenges, particularly when these models are deployed across diverse and high-volume contexts~\cite{deng2024deconstructing}. Ensuring consistent application of bias detection, privacy protections, and transparency measures across various domains is a complex endeavor. Bias mitigation, for instance, often requires specialized expertise and substantial resources, which may not be accessible to all organizations, leading to uneven ethical standards across AI applications~\cite{khan2022ethics}.

Scalability is especially critical in high-risk sectors such as healthcare and finance, where ethical lapses can have serious consequences. The EU AI Act and NIST guidelines recognize the importance of scalable solutions in these contexts. However, translating high-level guidelines into practical, domain-specific applications remains a challenge, as detailed by Fjeld et al., who examine the limitations of current AI frameworks in meeting diverse operational requirements across sectors~\cite{fjeld2020principled}. 

One pressing issue is the need for tools and frameworks to adapt to the varied operational scales at which LLMs are deployed. Binns et al. highlight this challenge, noting that most bias mitigation strategies lack the flexibility to be scaled up effectively, often requiring case-specific customization that hampers broader application~\cite{binns2018fairness}. Furthermore, the resource-intensive nature of comprehensive ethical compliance poses additional barriers for smaller organizations, which may lack the funding to implement scalable solutions~\cite{hoffmann2020artificial, cath2018governing}. 

\begin{tcolorbox}[title=Recommendation]
    We recommend building scalability into the core guidelines and frameworks for large-scale LLMs, ensuring they support the creation of ethical AI solutions that are accessible and inclusive.
\end{tcolorbox}

\subsection{Engaging with End Users and interdisciplinary collaboration}

The ethical deployment of LLMs necessitates active engagement with end users and robust interdisciplinary collaboration. Historically, AI development has often prioritized technical performance over societal impact, leading to unintended consequences such as bias and reduced trust among users~\cite{murphy2021artificial}. Involving end users, especially those from marginalized communities disproportionately affected by AI biases, is crucial for aligning LLMs with societal values and human rights~\cite{hagerty2019global}. 

Interdisciplinary collaboration is equally vital in the ethical implementation of LLMs. Bringing together technical experts, legal professionals, ethicists, and domain specialists facilitates the integration of ethical considerations into technical processes~\cite{pink2024trust}. Such collaborations have been initiated by organizations like the European Union and IEEE, setting a precedent for effective cooperation~\cite{al2024ethical}. 
However, challenges persist in achieving meaningful engagement and collaboration. Power imbalances, differing priorities, and communication barriers can hinder effective participation~\cite{keles2023navigating}. Addressing these challenges requires deliberate efforts to create inclusive environments where diverse perspectives are valued and integrated into the AI development process~\cite{gianni2022governance}.

\begin{tcolorbox}[title=Recommendation,breakable]
    \textcolor{black}{We encourage LLM developers to collaborate directly with representatives of marginalized communities - groups that have historically experienced social, economic, or political disadvantages and are often under-represented in AI design (racial or ethnic minorities, people with disabilities, LGBTQ+ individuals, indigenous people) through design workshops, community advisory boards, iterative user testing sessions. Such engagements can surface context specific biases and validate the system's outputs meet diverse needs and values.}
\end{tcolorbox}

\section{Threats to Validity}~\label{sec:threats}

In this section, we discuss the threats to validity of our study.

\textbf{Internal Validity:} One of the key threats to internal validity in a systematic mapping study (SMS) is selection bias, which can arise from subjective interpretation during the study selection process. To mitigate this, we employed multiple strategies. We identified a set of keywords considered relevant to the ethical use of LLMs, tested them, and consulted a university librarian \textcolor{black}{with expertise in systematic review search design, who recommended adding controlled vocabulary terms} to refine the keywords. We conducted searches across six databases to ensure a broader variety of studies could be included. Pilot tests were performed and validated by all authors to ensure the reliability of the results. Additionally, forward snowballing was performed to capture studies that may not have been initially identified.

\textbf{Construct Validity:}
Construct validity in our study refers to the extent to which the selected studies are relevant and appropriate to our research goals. To address this, we selected papers that directly aligned with our research questions (RQs) and excluded papers that focused solely on LLMs without discussing the ethical issues associated with their use or deployment. We also held meetings and discussions to establish the inclusion and exclusion criteria for the selected papers. The timeframe for our SMS was set between 2023 and July 2024, so any study published after July 2024 would not be reflected in our results.

\textcolor{black}{\textbf{External Validity:} Our mapping draws primarily on studies originating in Western jurisdictions, each of which has its own cultural norms and regulatory frameworks. As a result, the universal ethical dimensions and mitigation strategies we identify may not translate directly to non-Western settings with different legal requirements or value systems. We intend to incorporate empirical work and guidelines from diverse regions, such as Asia, Africa, and Latin America, into our future work to validate findings from different studies across various cultural and legal landscapes. Furthermore, by excluding papers under four pages, we might have omitted brief but potentially relevant contributions. However, this criterion helped us focus on studies with sufficient methodological and conceptual content. We would focus on revisiting these studies in our future work for more information.}

\textbf{Conclusion Validity:}
One of the major threats to conclusion validity in systematic mapping studies is bias in data extraction. In our study, the data extraction process was guided by our RQs, ensuring that the selected data was directly relevant to our study objectives. To mitigate potential bias, we used Google Forms to facilitate the coding process, enabling us to categorize data systematically through thematic analysis. This approach allowed us to group the findings based on predefined codes and themes derived from existing literature. As new themes emerged during the coding process, they were incorporated as needed. We held regular meetings to discuss and refine the data extraction and analysis processes, ensuring agreement on selecting relevant data and how it should be presented. Another concern regarding validity is the quality of evidence and the potential for publication bias. We acknowledge that a majority of our included studies (26 of 39) are conceptual, which may introduce author perspective bias into the reported strategies and skew the prominence of specific ethical dimensions. Moreover, our synthesis may be affected by publication bias, since positive or novel conceptual contributions are more likely to appear in the literature. To mitigate this, we searched across multiple databases and manually checked reference lists; we will also incorporate empirical investigations to validate the results from the conceptual studies.
\section{Conclusion}~\label{sec:conclusion}

In conclusion, this study offers a comprehensive examination of ethical concerns surrounding LLMs, categorized into five primary dimensions—safety, transparency, accountability, privacy, and bias—developed from a synthesis of selected guidelines and existing literature. Within these dimensions, numerous ethical themes emerge, highlighting complex issues such as the potential for LLMs to propagate misinformation, the opacity in model decision-making, challenges in establishing accountability, risks to user privacy, and the perpetuation of social biases.
Our findings reveal that, although numerous mitigation strategies have been proposed to address these dimensions, substantial implementation challenges remain due to technical complexities, evolving regulatory frameworks, ethical dilemmas, and user awareness and data quality issues. These obstacles highlight the need for adaptable, interdisciplinary ethical frameworks/guidelines capable of evolving alongside AI advancements. By elucidating these ethical dimensions and the practical difficulties in deploying mitigation strategies, this study contributes valuable insights to the discourse on responsible AI, informing future research, policy, and best practices to align LLM development with ethical standards better.

\section*{Acknowledgements}
Prof John Grundy is supported by Australian Research Council (ARC) under the Future Fellowship grant number FL190100035. The authors thank our librarian (wishing to remain anonymous) for her expert assistance in designing and validating our search strategy.

\bibliographystyle{elsarticle-num}
\bibliography{paper}

\section{Appendix A: Selected Primary Studies}
\begin{enumerate}[label=P\arabic*]

    \item \textcolor{blue}{\label{P1}}: Sathe, N., Deodhe, V., Sharma, Y., \& Shinde, A. (2023, December). A Comprehensive Review of AI in Healthcare: Exploring Neural Networks in Medical Imaging, LLM-Based Interactive Response Systems, NLP-Based EHR Systems, Ethics, and Beyond. In 2023 International Conference on Advanced Computing \& Communication Technologies (ICACCTech) (pp. 633-640). IEEE.
    
    \item \label{P2}: Ruane, E., Birhane, A., \& Ventresque, A. (2019). Conversational AI: Social and Ethical Considerations. AICS, 2563, 104-115.
    
    \item \label{P3}: Malic, V. Q., Kumari, A., \& Liu, X. (2023, December). Racial skew in fine-tuned legal AI language models. In 2023 IEEE International Conference on Data Mining Workshops (ICDMW) (pp. 245-252). IEEE.
    
    \item \label{P4}: Bansal, R. (2022). A survey on bias and fairness in natural language processing. arXiv preprint arXiv:2204.09591.
    
    \item \label{P5}: Bang, J., Lee, B. T., \& Park, P. (2023, August). Examination of ethical principles for llm-based recommendations in conversational ai. In 2023 International Conference on Platform Technology and Service (PlatCon) (pp. 109-113). IEEE.
    
    \item \label{P6}: Lofstead, J. (2023, September). Economic, Societal, Legal, and Ethical Considerations for Large Language Models. In 2023 Fifth International Conference on Transdisciplinary AI (TransAI) (pp. 155-162). IEEE.
    
    \item \label{P7}: Khoury, R., Avila, A. R., Brunelle, J., \& Camara, B. M. (2023, October). How secure is code generated by chatgpt?. In 2023 IEEE International Conference on Systems, Man, and Cybernetics (SMC) (pp. 2445-2451). IEEE.
    
    \item \label{P8}: Jiao, J., Afroogh, S., Xu, Y., \& Phillips, C. (2024). Navigating llm ethics: Advancements, challenges, and future directions. arXiv preprint arXiv:2406.18841.
    
    \item \label{P9}: Cai, Z., Chang, X., \& Li, P. (2023, November). HCPP: A Data-Oriented Framework to Preserve Privacy during Interactions with Healthcare Chatbot. In 2023 IEEE International Performance, Computing, and Communications Conference (IPCCC) (pp. 283-290). IEEE.
    
    \item \label{P10}: Piñeiro-Martín, A., García-Mateo, C., Docío-Fernández, L., \& Lopez-Perez, M. D. C. (2023). Ethical challenges in the development of virtual assistants powered by large language models. Electronics, 12(14), 3170.
    
    \item \label{P11}: Parray, A. A., Inam, Z. M., Ramonfaur, D., Haider, S. S., Mistry, S. K., \& Pandya, A. K. (2023). ChatGPT and global public health: applications, challenges, ethical considerations and mitigation strategies.
    
    \item \label{P12}: Khan, M. S., \& Umer, H. (2024). ChatGPT in finance: Applications, challenges, and solutions. Heliyon, 10(2).
    
    \item \label{P13}: Harrer, S. (2023). Attention is not all you need: the complicated case of ethically using large language models in healthcare and medicine. EBioMedicine, 90.
    
    \item \label{P14}: Chauncey, S. A., \& McKenna, H. P. (2023). A framework and exemplars for ethical and responsible use of AI Chatbot technology to support teaching and learning. Computers and Education: Artificial Intelligence, 5, 100182.
    
    \item \label{P15}: Ansarullah, S. I., Kirmani, M. M., Alshmrany, S., \& Firdous, A. (2024). Ethical issues around artificial intelligence. In A Biologist s Guide to Artificial Intelligence (pp. 301-314). Academic Press.
    
    \item \label{P16}: Patton, D. U., Landau, A. Y., \& Mathiyazhagan, S. (2023). ChatGPT for social work science: Ethical challenges and opportunities. Journal of the Society for Social Work and Research, 14(3), 553-562.
    
    \item \label{P17}: Wu, X., Duan, R., \& Ni, J. (2024). Unveiling security, privacy, and ethical concerns of ChatGPT. Journal of Information and Intelligence, 2(2), 102-115.
    
    \item \label{P18}: Guo, D., Chen, H., Wu, R., \& Wang, Y. (2023). AIGC challenges and opportunities related to public safety: a case study of ChatGPT. Journal of Safety Science and Resilience, 4(4), 329-339.
    
    \item \label{P19}: Oviedo-Trespalacios, O., Peden, A. E., Cole-Hunter, T., Costantini, A., Haghani, M., Rod, J. E., ... \& Reniers, G. (2023). The risks of using ChatGPT to obtain common safety-related information and advice. Safety science, 167, 106244.
    
    \item \label{P20}: Head, C. B., Jasper, P., McConnachie, M., Raftree, L., \& Higdon, G. (2023). Large language model applications for evaluation: Opportunities and ethical implications. New directions for evaluation, 2023(178-179), 33-46.
    
    \item \label{P21}: MacIntyre, M. R., Cockerill, R. G., Mirza, O. F., \& Appel, J. M. (2023). Ethical considerations for the use of artificial intelligence in medical decision-making capacity assessments. Psychiatry research, 328, 115466.
    
    \item \label{P22}: Grote, T., \& Berens, P. (2024). A paradigm shift?—On the ethics of medical large language models. Bioethics, 38(5), 383-390.
    
    \item \label{P23}: Behnia, R., Ebrahimi, M. R., Pacheco, J., \& Padmanabhan, B. (2022, November). Ew-tune: A framework for privately fine-tuning large language models with differential privacy. In 2022 IEEE International Conference on Data Mining Workshops (ICDMW) (pp. 560-566). IEEE.
    
    \item \label{P24}: Wang, C., Liu, S., Yang, H., Guo, J., Wu, Y., \& Liu, J. (2023). Ethical considerations of using ChatGPT in health care. Journal of Medical Internet Research, 25, e48009.
    
    \item \label{P25}: Mitsunaga, T. (2023, October). Heuristic Analysis for Security, Privacy and Bias of Text Generative AI: GhatGPT-3.5 case as of June 2023. In 2023 IEEE International Conference on Computing (ICOCO) (pp. 301-305). IEEE.
    
    \item \label{P26}: Gan, W., Qi, Z., Wu, J., \& Lin, J. C. W. (2023, December). Large language models in education: Vision and opportunities. In 2023 IEEE international conference on big data (BigData) (pp. 4776-4785). IEEE.
    
    \item \label{P27}: Kshetri, N. (2023). Cybercrime and privacy threats of large language models. IT Professional, 25(3), 9-13.
    
    \item \label{P28}: Zhuo, T. Y., Huang, Y., Chen, C., \& Xing, Z. (2023). Red teaming chatgpt via jailbreaking: Bias, robustness, reliability and toxicity. arXiv preprint arXiv:2301.12867.
    
    \item \label{P29}: Zhang, X., Xu, H., Ba, Z., Wang, Z., Hong, Y., Liu, J., ... \& Ren, K. (2024). Privacyasst: Safeguarding user privacy in tool-using large language model agents. IEEE Transactions on Dependable and Secure Computing.
    
    \item \label{P30}: Mökander, J., Schuett, J., Kirk, H. R., \& Floridi, L. (2023). Auditing large language models: a three-layered approach. AI and Ethics, 1-31.
    
    \item \label{P31}: Curzon, J., Kosa, T. A., Akalu, R., \& El-Khatib, K. (2021). Privacy and artificial intelligence. IEEE Transactions on Artificial Intelligence, 2(2), 96-108.
    
    \item \label{P32}: Bano, M., Hoda, R., Zowghi, D., \& Treude, C. (2024). Large language models for qualitative research in software engineering: exploring opportunities and challenges. Automated Software Engineering, 31(1), 8.
    
    \item \label{P33}: Khoje, M. (2024, February). Navigating Data Privacy and Analytics: The Role of Large Language Models in Masking conversational data in data platforms. In 2024 IEEE 3rd International Conference on AI in Cybersecurity (ICAIC) (pp. 1-5). IEEE.
    
    \item \label{P34}: Jeyaraman, M., Balaji, S., Jeyaraman, N., \& Yadav, S. (2023). Unraveling the ethical enigma: artificial intelligence in healthcare. Cureus, 15(8).
    
    \item \label{P35}: Ferrara, E. (2023). Should chatgpt be biased? challenges and risks of bias in large language models. arXiv preprint arXiv:2304.03738.
    
    \item \label{P36}: Akinci D'Antonoli, T., Stanzione, A., Bluethgen, C., Vernuccio, F., Ugga, L., Klontzas, M. E., ... \& Koçak, B. (2023). Large language models in radiology: fundamentals, applications, ethical considerations, risks, and future directions. Diagnostic and Interventional Radiology, Epub-ahead.
    
    \item \label{P37}: Belzner, L., Gabor, T., \& Wirsing, M. (2023, October). Large language model assisted software engineering: prospects, challenges, and a case study. In International Conference on Bridging the Gap between AI and Reality (pp. 355-374). Cham: Springer Nature Switzerland.
    
    \item \label{P38}: Liyanage, U. P., \& Ranaweera, N. D. (2023). Ethical considerations and potential risks in the deployment of large language models in diverse societal contexts. Journal of Computational Social Dynamics, 8(11), 15-25.
    
    \item \label{P39}: Hao, J., von Davier, A. A., Yaneva, V., Lottridge, S., von Davier, M., \& Harris, D. J. (2024). Transforming assessment: The impacts and implications of large language models and generative ai. Educational Measurement: Issues and Practice, 43(2), 16-29.
\end{enumerate}

\clearpage
\section{Appendix B: Database Search Strings}
\label{AppendixB}

\begin{center}
  \scriptsize
  \setlength{\tabcolsep}{3pt}%
  \framebox[\linewidth]{%
    \begin{tabular}{|p{0.2\linewidth}|p{0.75\linewidth}|}
      \hline
      \textbf{Database} & \textbf{Search String} \\
      \hline
      IEEE &
      ((``Large Language Model*'' OR LLMs)
      AND (guideline* OR standard* OR framework* OR compliance OR principles OR practices OR governance OR impact OR oversight OR algorithmic OR policy OR policies)
      AND (development OR deployment OR use OR design OR implementation)
      AND (ethics OR ethical OR moral OR bias OR fairness OR transparency OR accountability OR privacy OR security OR sustainability OR responsible OR trustworthiness OR equit* OR inclus* OR diversity OR legal OR rights OR cultural)) \\
      \hline
      ACM DL &
      (([All: ``large language model*''] OR [All: llms])
      AND ([All: guideline*] OR [All: standard*] OR [All: framework*] OR [All: compliance] OR [All: principles] OR [All: practices] OR [All: governance] OR [All: impact] OR [All: oversight] OR [All: algorithmic] OR [All: policy] OR [All: policies])
      AND ([All: development] OR [All: deployment] OR [All: use] OR [All: design] OR [All: implementation])
      AND ([All: ethics] OR [All: ethical] OR [All: moral] OR [All: bias] OR [All: fairness] OR [All: transparency] OR [All: accountability] OR [All: privacy] OR [All: security] OR [All: sustainability] OR [All: responsible] OR [All: trustworthiness] OR [All: equit*] OR [All: inclus*] OR [All: diversity] OR [All: legal] OR [All: rights] OR [All: cultural])) \\
      \hline
      ProQuest &
      ((noft(``Large Language Model*'') OR noft(LLMs))
      AND (noft(guideline*) OR noft(standard*) OR noft(framework*) OR noft(compliance) OR noft(principles) OR noft(practices) OR noft(governance) OR noft(impact) OR noft(oversight) OR noft(algorithmic) OR noft(policy) OR noft(policies))
      AND (noft(development) OR noft(deployment) OR noft(use) OR noft(design) OR noft(implementation))
      AND (noft(ethics) OR noft(ethical) OR noft(moral) OR noft(bias) OR noft(fairness) OR noft(transparency) OR noft(accountability) OR noft(privacy) OR noft(security) OR noft(sustainability) OR noft(responsible) OR noft(trustworthiness) OR noft(equit*) OR noft(inclus*) OR noft(diversity) OR noft(legal) OR noft(rights) OR noft(cultural))) \\
      \hline
      Web of Science &
      (("Large Language Model*" OR llms)
      AND (guideline* OR standard* OR framework* OR compliance OR principles OR practices OR governance OR impact OR oversight OR algorithmic OR policy OR policies)
      AND (development OR deployment OR use OR design OR implementation)
      AND (ethics OR ethical OR moral OR bias OR fairness OR transparency OR accountability OR privacy OR security OR sustainability OR responsible OR trustworthiness OR equit* OR inclus* OR diversity OR legal OR rights OR cultural)) \\
      \hline
      Wiley Online Library &
      (("Large Language Model*" OR llms)
      AND (guideline* OR standard* OR framework* OR compliance OR principles OR practices OR governance OR impact OR oversight OR algorithmic OR policy OR policies)
      AND (development OR deployment OR use OR design OR implementation)
      AND (ethics OR ethical OR moral OR bias OR fairness OR transparency OR accountability OR privacy OR security OR sustainability OR responsible OR trustworthiness OR equit* OR inclus* OR diversity OR legal OR rights OR cultural)) \\
      \hline
      Science Direct &
      (``Large Language Model'' OR ``LLMs'')
      AND (``guideline'' OR ``standard'')
      AND (``development'' OR ``design'')
      AND (``ethics'' OR ``moral'') \\
      \hline
    \end{tabular}%
  }
\end{center}

\end{document}